\documentclass[12pt]{iopart}
\usepackage{url}
\usepackage{multirow}
\usepackage{graphicx}
\usepackage{iopams}
\usepackage{hhline}
\usepackage{epsfig}
\usepackage{epstopdf}
\usepackage{booktabs}
\usepackage{multirow}
\usepackage{siunitx}
\usepackage{adjustbox}
\usepackage{caption}
\usepackage{array}
\usepackage{cite}
\usepackage{etoolbox}
\usepackage{graphicx, nicefrac}
\newcolumntype{L}[1]{>{\raggedright\let\newline\\\arraybackslash\hspace{0pt}}m{#1}}
\newcolumntype{C}[1]{>{\centering\let\newline\\\arraybackslash\hspace{0pt}}m{#1}}
\newcolumntype{R}[1]{>{\raggedleft\let\newline\\\arraybackslash\hspace{0pt}}m{#1}}

\begin{document}
\title[Validity of BA Hypothesis for  heavy nuclei]
{Validity of Brink-Axel Hypothesis for calculations of allowed stellar weak rates of heavy nuclei}

\author{Fakeha Farooq$^1$, Jameel-Un Nabi$^{2,3}$, Ramoona Shehzadi$^1$}
\address{$^1$ Department of Physics, University of the Punjab,
Lahore, Pakistan.}
\address{$^2$ University of Wah, Quaid Avenue, Wah Cantt 47040, Punjab, Pakistan.}
\address{$^3$ Faculty of Engineering Sciences, GIK Institute of Engineering Sciences and Technology, Topi
23640, Khyber Pakhtunkhwa, Pakistan.}

\ead{ramoona.physics@pu.edu.pk}

\begin{abstract}
The knowledge of beta-decay transitional probabilities and Gamow-Teller (GT) strength functions from highly excited states
of nuclides is of particular importance for applications to astrophysical network calculations of nucleosynthesis
in explosive stellar events. These quantities  are challenging to achieve from measurements or computations using various nuclear
models. Due to unavailability of feasible alternatives, many theoretical studies  often rely on the Brink-Axel (BA) hypothesis, that
is, the response of strength functions depends merely on the transition energy of the parent nuclear ground state and is
independent of the underlying details of the parent state, for the calculation of stellar rates. BA hypothesis has been used in many applications from nuclear
structure determination to nucleosynthesis yield in the astrophysical matter. We explore here the the validity of BA hypothesis
in the calculation of stellar beta-decay (BD) and electron capture (EC) weak rates of fp- and fpg-shell nuclides for GT transitions.
Strength functions have been computed employing the fully
microscopic proton-neutron QRPA (quasi-particle random-phase approximation) within a broad density,
$\rho$Y$_{e}$ = (10 - 10$^{11}$) [g\;cm$^{-3}$], and temperature, T = (1$-$30) [GK], grid relevant to the
pre-collapse astrophysical environment. Our work provides evidence that the use of the approximation based on the BA hypothesis
does not lead to  reliable calculations of excited states strength functions under extreme temperature-density conditions characteristic of
presupernova and supernova evolution of massive stars. Weak rates obtained by incorporating the BA hypothesis in the calculation of strength functions
substantially deviate from the rates based on the  state-by-state microscopically calculated strength functions. Deviation in the two calculations becomes significant as early as neon burning phases of massive stars.
The deviation in the calculation of BD rates is even more pronounced, reaching up to three orders of magnitude.

\end{abstract}

%\pacs{23.40.Bw, 23.40.-s, 26.30.Jk, 26.50.+x, 97.10.Cv} \vspace{2pc}
\noindent{\it Keywords}: Brink-Axel hypothesis, proton neutron QRPA model, Beta decay, Electron capture, GT transitions

\submitto{\PS}
\maketitle
%\ioptwocol
%
As is well understood, leptonic abundance in proportion to baryon, the entropy of the inner stellar matter and the size of
collapsing core are the fundamental quantities involved in the dynamics of core-collapse of massive stars. Over the
last phases, in the life journey of a dying massive star with M $\gtrsim$ 10M$_{\odot}$, a substantial impact on
these quantities is owing to the weak nuclear processes including beta-decay (BD) and electron capture (EC). For stellar
densities $\leq$ 10$^{11}$ g\;cm$^{-3}$, these reactions are mainly governed by Gamow-Teller (GT)
transitions~\cite{Bet90,Iwa99,Hix03}. During the collapse, density and temperature of the inner stellar
environment approach  10$^{10}$\;g\;cm$^{-3}$ and 1\;GK,  respectively. These physical conditions initiate EC on free protons and the
protons inside heavy nuclides~\cite{Cha39,Bro82}. This process lowers the electron-to-baryon number (Y$_{e}$)
and shifts the stellar nuclear matter to high-mass nuclides having larger neutron excess. In the neutron-rich environment, the
BD reaction energies (Q$_{\beta}$) become dominant and make BD considerably more favorable and a competing process over EC. In
addition, abundantly produced (anti-) neutrinos during the weak reactions (and other processes) are responsible for the
variation in the core entropy on a large scale~\cite{Lan00,Ful91,Heg01}. The EC (BD) processes profoundly
influence the initial dynamics of core-collapse by changing the entropy, Y$_{e}$ and the size of collapsing core via M$_{ch}$.
Consequently, it triggers the supernovae explosions and alters the nucleosynthesis yield of core-collapse and
explosive burning phases from the O-Ne-Mg core of white dwarf~\cite{Nab07} to pre-supernovae cores of high mass stars~\cite{Heg01,Ful82,Auf94,Nab21}.

Understanding of these processes  demands reliable calculation of GT strength functions or transitional probabilities of nuclear weak transitions.
Extreme hot and dense physical conditions during stellar evolution, lead to highly populated excited states of decaying
and residual nuclides. Such excited- to-excited state transitions have significant contribution to the overall weak
rate values~\cite{Nab04}. However, it is much more difficult to access such transitions either from the measurements or
through theory. For example, the (n,p) and (p,n) type reactions measurements~\cite{Doe75,Goo80} and the interacting
shell-model based calculations~\cite{Cau05} provide data of transitional strengths merely from the ground and low-lying
excited states. The astrophysical calculations require transitions from high-lying states up to or even more than
hundreds of states.
Thus, due to a lack of practical alternatives, the most appealing resolution is the implementation
of generalized BA hypothesis~\cite{Ful82} for GT transition
strength functions.

Initially, the idea that the giant electric dipole resonances of excited states should remain at the same energy relative
to the ground state resonance, with the argument that the giant electric dipole resonance is insensitive to the
details of the initial state, was hypothesized by Brink~\cite{Bri55} and, independently by Axel~\cite{Axe62}.
This assumption is known as Brink-Axel (BA) hypothesis. Later, the generalized form of this hypothesis has been
widely employed, e.g., in the calculations of magnetic-dipole (M1) transitions~\cite{Loe12}, and for GT and Fermi
transitional strengths~\cite{Lan00} in BD and EC studies. In the first place, Fuller, Fowler, and Newman
(FFN)~\cite{Ful82,Fuller} adopted the BA hypothesis for use in the GT resonances of excited states to approximate the
centroid energies. Their work is considered as a foundation in the field of astrophysics.
Afterwards, the FFN approach (with modifications) has been
widely implemented to calculate the weak rates~\cite{Auf94,Kar94,Lan00,Kaj88,Fis13,Cau99,Mar99,Col12}.
In Independent-particle~\cite{Fuller} and Large-scale shell~\cite{Lan00} models, the use
of the BA hypothesis for excited states was augmented by back resonances (see~\cite{Nab22} for detail).
However, in the configuration-interaction shell model studies~\cite{Joh15,Lu18}, the authors reported that
the energy-weighed (non-energy-weighed) sum rules evolve with increasing initial energy in a smooth manner for
E1, E2, M1 and GT transitions, leading to the violation of the BA hypothesis for GT strengths.
Some works show that sum rules are independent of initial energy for E1~\cite{Kru19} and GT transitions~\cite{Lan00} for
a few lowest energy transitions. In many other experimental and theoretical studies, the reliability of the BA hypothesis
has remained controversial. Such as for Gamma ray strength functions, several experimental studies
(e.g.,~\cite{Ram81,Sze83,Gut16}) are in support of it and some report failure of the BA hypothesis
(e.g.,~\cite{Ang12}). The study done by Larsen \etal~\cite{Lar07} for the calculations of M1, E1 and E2, and the studies
of~\cite{Fra97,Mis14} for GT transitions have questioned the reliability of the BA hypothesis.
The results of many other studies, employing the famous proton-neutron QRPA (quasi-particle random phase approximation)
theory, do not agree with the reliability of the use of the BA hypothesis for astrophysical applications.
The reported weak rates of sd-~\cite{Nab07,Nab99}, fp- and fpg-shell
nuclides~\cite{Nab04,Nab09,Maj16,Nab12,Nab19} in earlier proton-neutron QRPA studies were
orders of magnitude different than those based on the BA hypothesis at astrophysical
conditions. Furthermore, the studies~\cite{Dzh10,Yuk20}, based on the finite-temperature proton-neutron QRPA model, violate
the BA hypothesis by showing that GT strengths evolve with temperature as the pairing interaction become weaker.
Recently, Herrera \etal~\cite{Her22} introduced a new
and modified version of generalized BA hypothesis named as {\it "energy-localized"} or simply {\it "local"}
BA hypothesis. In this study, the authors showed that  GT transitions of fp-shell nuclei follow the {\it "local"}
BA hypothesis, by using the interacting shell-model.

We have studied here the validity of the BA hypothesis in the calculation of stellar weak rates. Our work is a follow-up of our study performed earlier for sd-shell nuclides~\cite{Nab22}. For this purpose, we have picked the top 80 fp- and fpg-shell
nuclides including both BD and EC, related to the post Si burning phases of pre-supernova stars~\cite{Heg01,Auf94,Nab21}. According to a recent simulation study of presupernova evolution \cite{Nab21}, the authors proposed several new BD (EC) nuclides which were not present in the
independent particle model (IPM) compiled list.
The proton-neutron QRPA model  has been extensively used  by Nabi and  collaborators for the calculations of weak rates of GT~\cite{Nab07,Nab04,Nab12,Nab19} transitions. This model calculates GT strength functions off parent excited states in a fully microscopic state-to-state fashion. A large model space of 7 major oscillatory shells is accessible in this model. Therefore, arbitrarily heavy nuclide can be accommodated by the proton-neutron QRPA model. However, our model is not based on a thermal phonon vacuum in which the temperature effect is included in the mean-field. There are studies based on QRPA formalism that include the effect of finite temperature in the calculation of nuclear matrix elements (NMEs) of Fermi and GT transitions. These models are commonly referred to as finite-temperature QRPA (FTQRPA) and thermal QRPA (TQRPA). The studies performed by Dzhioev and collaborators are based on TQRPA approach in which temperature is taken into account within the thermo-field dynamics (TFD) formalism by using the Woods-Saxon potential, BCS pairing interactions and separable multipole and spin-multipole particle-hole interactions~\cite{Dzh10,Dzh2019}. Furthermore, a self-consistent FTQRPA approach within relativistic~\cite{Yuk20,Niu11} and non-relativistic~\cite{PaarFan} frameworks, based on nuclear energy density functionals for evaluation of nuclear weak interaction rates, is also available at finite temperature.
The current pn-QRPA model may be considered as a zero-temperature approximation for the calculation of NMEs. It is to be noted that our model calculates state-by-state rates in a microscopic fashion and roughly covers 15 MeV excitation energy in parent and daughter nuclei. This is a big advantage of the model. In trade-off we are forced to give up the notion of thermal phonon vacuum. Furthermore, the parent excited states, in the current model, are assumed to be populated using the Boltzmann distribution which is temperature dependent. This recipe is used frequently in previous stellar rate
calculations (e.g., independent particle model~\cite{Auf94,Fuller,Pru03}, full shell-model~\cite{Oda94} and Large-scale
shell-model~\cite{Lan00}). All these recipes, including the current model, incorporate the finite-temperature effect in phase-space calculations and not in NMEs. 

The outline of the paper is as follows. In Section~\ref{sec:formalism}, the proton-neutron QRPA formalism has been described.  Our results on the validity of the BA hypothesis are discussed in Section~\ref{sec:results}. Lastly in Section~\ref{sec:conclusions}, conclusion is presented.

\section{Theoretical Formalism}
\label{sec:formalism}
In this section, we briefly describe the frame-work of the proton-neutron QRPA theory, and the parameters used in the model to  calculate the astrophysical weak transition rates. This formalism has been adopted in many earlier studies
(e.g.,~\cite{Suh98}). In the  proton-neutron QRPA theory, the ground state is a vacuum for QRPA phonon, $\hat{\Gamma}_{\omega}|$QRPA$> = 0$, with phonon creation operator defined by

\begin{eqnarray}
\hat{\Gamma}^{\dagger}_{\omega}(\mu)&=& \sum_{\pi,\nu}X^{\pi\nu}_{\omega}(\mu)\hat{a}^{\dagger}_{\pi}\hat{a}^{\dagger}_{\bar{\nu}}-Y^{\pi\nu}_{\omega}(\mu)\hat{a}_{\nu}\hat{a}_{\bar{\pi}},
\label{Eq:PCO}
\end{eqnarray}
where $\nu$ and $\pi$, respectively, denote the neutron and proton single quasi-particle states, with $(\hat{a}^{\dagger}, \hat{a})$ are creation and annihilation operators of these states. The sum runs over all possible
$\pi\nu$-pairs which satisfy $\mu=m_{\pi}-m_{\nu}$ with $m_{\pi}(m_{\nu})$ being the third component of angular momentum.
The forward-going ($X_{\omega}$) and backward-going ($Y_{\omega}$) amplitudes, and energy ($\omega$) are, respectively, the eigenvector and eigenvalues of the
famous (Q)RPA equation
\begin{center}
\begin{equation}
  \left[ {\begin{array}{cc}
    M & N \\
  -N & -M \\
  \end{array} } \right]
   \left[ {\begin{array}{c}
    X \\
    Y\\
  \end{array} } \right] =  \omega \left[ {\begin{array}{c}
    X\\
    Y\\
  \end{array} } \right].
  \label{Eq:RPAM}
\end{equation}
\end{center}

\begin{figure}
\begin{center}
\includegraphics[width=0.8\textwidth]{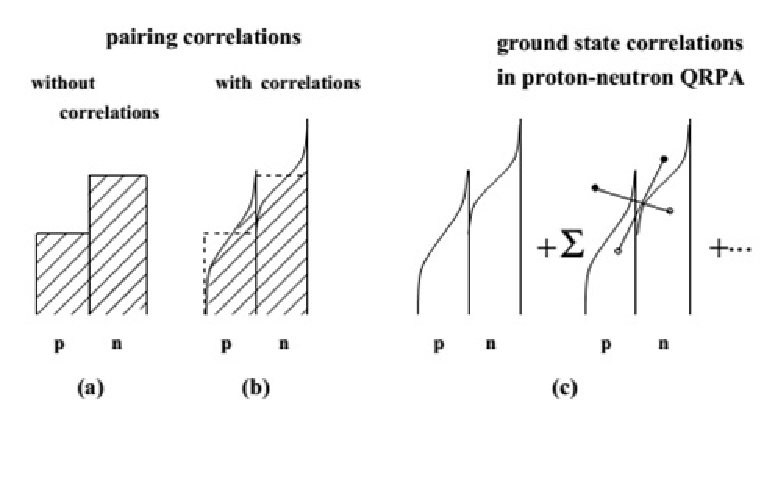}
\end{center}
\caption{\small Distributions of nucleons among single-particle orbits in a nucleus; (a) without pairing correlations (the simplest shell model), (b) with pairing correlations. (c) Ground state wave function in the proton-neutron quasi-particle RPA. The line connecting circles, which denotes quasi-particles, indicates angular momentum coupling of a proton-neutron pair. Both pairs have the same spin-parity.
}\label{figure1}
\end{figure}

The solution of RPA equation is obtained for each projection value $\mu = 0,\pm1$. Here, M and N matrices are given as
\begin{eqnarray}
M_{\pi\nu,\pi^{\prime}\nu^{\prime}}=&\delta_{\pi\nu,\pi^{\prime}\nu^{\prime}}(\varepsilon_{\pi}+\varepsilon_{\nu})\nonumber \\
&+V^{pp}_{\pi\nu,\pi^{\prime}\nu^{\prime}}(v_{\pi}v_{\nu}v_{\pi^{\prime}}v_{\nu^{\prime}}+u_{\pi}u_{\nu}u_{\pi^{\prime}}u_{\nu^{\prime}})\nonumber \\
&+V^{ph}_{\pi\nu,\pi^{\prime}\nu^{\prime}}(v_{\pi}u_{\nu}v_{\pi^{\prime}}u_{\nu^{\prime}}+u_{\pi}v_{\nu}u_{\pi^{\prime}}v_{\nu^{\prime}}),
\label{Mentry}
\end{eqnarray}
\begin{eqnarray}
N_{\pi\nu,\pi^{\prime}\nu^{\prime}}=&V^{pp}_{\pi\nu,\pi^{\prime}\nu^{\prime}}(u_{\pi}u_{\nu}v_{\pi^{\prime}}v_{\nu^{\prime}}+v_{\pi}v_{\nu}u_{\pi^{\prime}}u_{\nu^{\prime}})\nonumber \\
&-V^{ph}_{\pi\nu,\pi^{\prime}\nu^{\prime}}(v_{\pi}u_{\nu}u_{\pi^{\prime}}v_{\nu^{\prime}}+u_{\pi}v_{\nu}v_{\pi^{\prime}}u_{\nu^{\prime}}),
\label{Nentry}
\end{eqnarray}
where the quasi-particle energies ($\varepsilon_{\pi}, \varepsilon_{\nu}$) and the occupation
amplitudes ($u_{\pi(\nu)},v_{\pi(\nu)}$), which satisfy $u^{2}+v^{2} = 1$, are obtained from BCS calculations.
In the first step, in the framework of QRPA calculation, quasi-particle basis are constructed in terms of nucleon states and
defined by Bogoliubov transformation with pairing correlations. Figure~\ref{figure1} compares the nucleon distributions among the orbits for the cases without and with pairing correlations. When there are no correlations (Figure~\ref{figure1}(a)), low-lying orbits are completely filled, and high-lying orbits are empty (one orbit nearest to the Fermi energy can be partly filled). On the other hand, the pairing interaction smears out the distribution of nucleons (Figure~\ref{figure1}(b)). All nucleons are paired to J$_{\pi}$ = 0$^{+}$. This state is the vacuum for quasi-particles. The ground state wave function is illustrated in Figure~\ref{figure1}(c). The main component is the BCS ground state with no quasi-particles (Figure~\ref{figure1}(b)), and the leading admixtures are four-quasi-particle states. Secondly, in the quasi-particle proton-neutron pairs,
the computation of the RPA equation (\ref{Eq:RPAM}) is performed with separable GT residual forces, namely;
particle-hole (ph) and particle-particle (pp) forces. We take pp GT force ($\hat{V}_{pp(GT)}$) as~\cite{Kuz88}
\begin{equation}
\hat{V}_{pp(GT)} = -2\kappa_{GT}\sum_{\mu} (-1)^{\mu}\hat{P}^{\dagger}_{\mu}\hat{P}_{-\mu},
\label{ppGT}
\end{equation}
where
\begin{equation}
\hat{P}^{\dagger}_{\mu} = \sum_{j_{\pi}m_{\pi}{j_{\nu}m_{\nu}}} \langle j_{\nu}m_{\nu}|(\sigma_{\mu}\tau_{-})^{\dagger}|j_{\pi}m_{\pi} \rangle (-1)^{l_{\nu}+j_{\nu}-m_{\nu}}\hat{c}^{\dagger}_{j_{\pi}m_{\pi}}\hat{c}^{\dagger}_{j_{\nu}-m_{\nu}},
\end{equation}
and ph GT force ($\hat{V}_{ph(GT)}$) as~\cite{Halb67}
\begin{equation}
\hat{V}_{ph(GT)} = 2\chi_{GT}\sum_{\mu} (-1)^{\mu}\hat{R}_{\mu}\hat{R}^{\dagger}_{-\mu},
\label{phGT}
\end{equation}
where
\begin{equation}
\hat{R}_{\mu} = \sum_{j_{\pi}m_{\pi}{j_{\nu}m_{\nu}}} \langle j_{\pi}m_{\pi}|\sigma_{\mu}\tau_{-}|j_{\nu}m_{\nu} \rangle\hat{c}^{\dagger}_{j_{\pi}m_{\pi}}\hat{c}_{j_{\nu}m_{\nu}}.
\end{equation}

With positive values of force constants ($\chi_{GT}, \kappa_{GT}$), the pp and ph GT forces are, respectively, attractive and repulsive. With the use of the separable GT forces in the calculation, the RPA matrix equation reduces to a 4$^{th}$ order algebraic equation. The method to determine the roots of these equations is given in~\cite{Mut92}. This saves the computational time relative to the fully diagonalization of the
nuclear Hamiltonian.
In the RPA formalism, excitations from the ground state ($J^{\pi} = 0^+$) of an even-even nucleus is considered.
The ground-state of an odd-odd (odd-A) parent nucleus is expressed as a proton-neutron quasi-particle pair
(one quasi-particle) state of the smallest energy. Then two possible transitions are the phonon excitations (in which
quasi-particle merely plays the role of a spectator) and transition of quasi-particle itself. In
the latter case, correlations of phonon to the quasi-particle transitions are treated in first-order perturbation~\cite{Mol90,Ben88}. We next present quasiparticle transitions, construction of phonon-related multi-quasiparticle states (representing nuclear excited sates of even-even, odd-A and odd-odd nuclei) and formulae
of GT transitions within the current model using the recipe given by~\cite{Mut92}.
The phonon-correlated one quasi-particle states are defined by
\begin{eqnarray}
	|\pi_{corr}\rangle~=~a^\dagger_{\pi}|-\rangle +& \sum_{\nu, \omega}a^\dagger_{\nu}A^\dagger_{\omega}(\mu)|-\rangle \nonumber~\langle-|[a^\dagger_{\nu}A^\dagger_{\omega}(\mu)]^{\dagger}H_{31}a^\dagger_{\pi}|-\rangle \nonumber \\
	&\times E_{\pi}(\nu,\omega)
\label{opn1}
\end{eqnarray}
\begin{eqnarray}
	|\nu_{corr}\rangle~=~a^\dagger_{\nu}|-\rangle +& \sum_{\pi, \omega}a^\dagger_{\pi}A^\dagger_{\omega}(-\mu)|-\rangle \nonumber~\langle-|[a^\dagger_{\pi}A^\dagger_{\omega}(-\mu)]^{\dagger}H_{31}a^\dagger_{\nu}|-\rangle \nonumber \\
	&\times E_{\nu}(\pi,\omega)
	\label{opn2}
\end{eqnarray}
with
	\begin{equation}
		E_{a}(b,\omega)=\frac{1}{\epsilon_{a}-\epsilon_{b}-\omega}~~~~~~~a, b = \pi, \nu
		\label{opn3}
	\end{equation}
where the terms $E_{a}(b,\omega)$ can be modified to prevent the singularity in the transition amplitude
caused by the first-order perturbation of the odd-particle wave function. The first term in equations~(\ref{opn1}) and (\ref{opn2}) denotes the proton (neutron) quasi-particle state, while the second term denotes RPA correlated
phonons admixed with quasi-particle phonon coupled Hamiltonian H$_{31}$, which was accomplished by Bogoliubov
transformation from separable pp and ph GT interaction forces. The summation applies to all phonon states and
neutron (proton) quasi-particle states, satisfying $m_{\pi}-m_{\nu}=\mu$ with $\pi_{\pi}\pi_{\nu}=1$. Calculation of the quasi-particle transition amplitudes for correlated states
can be seen in~\cite{Mut89}. The amplitudes of GT transitions in terms of separable forces are
\begin{eqnarray}
		<{\pi_{corr}}|\tau_-\sigma_{\mu}|{\nu_{corr}}>~=~ q^U_{\pi\nu}+ 2\chi_{GT} [q^U_{\pi\nu}\sum\limits_{\omega}(Z^{-2}_\omega E_\pi(\nu,\omega)+Z^{+2}_{\omega}E_\nu(\pi,\omega)) \nonumber\\
		-q^V_{\pi\nu}\sum\limits_{\omega}Z^-_{\omega}Z^+_{\omega}(E_\pi(\nu,\omega)+E_\nu(\pi,\omega))]
		+2\kappa_{GT}[q_{\pi\nu}\sum\limits_{\omega}(Z^-_{\omega}Z^{--}_{\omega}E_\pi(\nu,\omega)-Z^+_{\omega}Z^{++}_{\omega}E_\nu(\pi,\omega)) \nonumber\\
		-\tilde{q}_{\pi\nu}\sum\limits_{\omega}(Z^-_{\omega}Z^{++}_{\omega}E_\pi(\nu,\omega)-Z^+_{\omega}Z^{--}_{\omega}E_\nu(\pi,\omega))],
\label{opn4}	
	\end{eqnarray}
	\begin{eqnarray}
		%	\begin{split}
			<{\pi_{corr}}|\tau_+\sigma_{\mu}|{\nu_{corr}}>=q^V_{\pi\nu}+2\chi_{GT}[q^V_{\pi\nu}\sum\limits_{\omega}(Z^{+2}_{\omega}E_\pi(\nu,\omega)+Z^{-2}_{\omega}E_\nu(\pi,\omega)) \nonumber\\
			-q^U_{\pi\nu}\sum\limits_{\omega}Z^-_{\omega}Z^+_{\omega}(E_\pi(\nu,\omega)+E_\nu(\pi,\omega))]
			+2\kappa_{GT}[\tilde{q}_{\pi\nu}\sum\limits_{\omega}(Z^+_{\omega}Z^{++}_{\omega}E_\pi(\nu,\omega)
			-Z^-_{\omega}Z^{--}_{\omega}E_\nu(\pi,\omega))\nonumber\\-q_{\pi\nu}\sum\limits_{\omega}(Z^+_{\omega}Z^{--}_{\omega}E_\pi(\nu,\omega)-Z^-_{\omega}Z^{++}_{\omega}E_\nu(\pi,\omega))],
			%	\end{split}
			\label{opn5}
	\end{eqnarray}
	
	\begin{equation}
		<{\nu_{corr}}|\tau_{\pm}\sigma_{-\mu}|{\pi_{corr}}>=(-1)^{\mu}<{\pi_{corr}}|\tau_{\mp}\sigma_{\mu}|{\nu_{corr}}>.
		\label{opn6}
	\end{equation}
In equations (\ref{opn4}), (\ref{opn5}) and (\ref{opn6}), $\sigma_{\mu}$ and $\tau_{\pm}$ are spin and iso-spin type operators,  respectively, and other symbols $q_{\pi\nu}$ ($\tilde{q}_{\pi\nu}$), $q^U_{\pi\nu}$ ($q^V_{\pi\nu}$), $Z^-_{\omega}$ ($Z^+_{\omega}$) and $Z^{--}_{\omega}$ ($Z^{++}_{\omega}$) are defined as
\begin{eqnarray}
q_{\pi\nu}=f_{\pi\nu}u_\pi v_\nu,~~~~ q_{\pi\nu}^{U}=f_{\pi\nu}u_\pi u_\nu, \nonumber \\
\tilde q_{\pi\nu}=f_{\pi\nu}v_\pi u_\nu,~~~~_{\pi\nu}^{V}=f_{\pi\nu}v_\pi v_\nu \nonumber \\
	Z^{-}_{\omega}= \sum_{\pi,\nu}(X^{\pi\nu}_{\omega}q_{\pi\nu}-Y^{\pi\nu}_{\omega}\tilde q_{\pi\nu}),\nonumber \\
	Z^{+}_{\omega}= \sum_{\pi,\nu}(X^{\pi\nu}_{\omega}\tilde q_{\pi\nu}-Y^{\pi\nu}_{\omega}q_{\pi\nu}),
\nonumber \\
	Z^{--}_{\omega}= \sum_{\pi,\nu}(X^{\pi\nu}_{\omega}q^{U}_{\pi\nu}+Y^{\pi\nu}_{\omega}q^{V}_{\pi\nu}),
\nonumber \\
	Z^{+ +}_{\omega}=
	\sum_{\pi\nu}(X^{\pi,\nu}_{\omega}q^{V}_{\pi\nu}+Y^{\pi\nu}_{\omega}q^{U}_{\pi\nu}).
\end{eqnarray}
The terms $X^{\pi\nu}_{\omega}$ and $Y^{\pi\nu}_{\omega}$ are defined earlier and other symbols have usual meanings.
The idea of quasi-particle transitions with first-order phonon correlations can be extended to an odd-odd parent nucleus.
The ground state is assumed to be a proton-neutron quasi-particle pair state of the smallest energy. The GT transitions
of the quasi-particle lead to two-proton or two-neutron quasi-particle states in the even-even daughter nucleus. The
two quasi-particle states are constructed with phonon correlations and given by
	\begin{eqnarray}
			|{\pi \nu_{corr}}>~=~a_\pi^\dagger a^\dagger_\nu|{-}>+\frac{1}{2}\sum\limits_{\pi'_1,\pi'_2,\omega}a^\dagger_{\pi'_1}a^\dagger_{\pi'_2}A^\dagger_{\omega}(-\mu)|{-}>~~~~~~~~~~~~~~~\nonumber\\\times <{-}|[a^\dagger_{\pi'_1}a^\dagger_{\pi'_2}A^\dagger_{\omega}(-\mu)]^\dagger H_{31}a^\dagger_\pi a^\dagger_\nu|{-}>E_{\pi\nu}(\pi'_1\pi'_2,\omega)+\frac{1}{2}\sum\limits_{\nu'_1,\nu'_2,\omega}a^\dagger_{\nu'_1}a_{\nu'_2}A^\dagger_{\omega}(\mu)|{-}>\nonumber\\\times <{-}|[a^\dagger_{\nu'_1}a^\dagger_{\nu'_2}A^\dagger_{\omega}(\mu)]^\dagger
			H_{31}a^\dagger_\pi a^\dagger_\nu|{-}>E_{\pi\nu}(\nu'_1\nu'_2,\omega),
			\label{opn7}
	\end{eqnarray}
	\begin{eqnarray}
		%	\begin{split}
			<{\pi_1\pi_{2corr}}|~=~a^\dagger_{\pi_1}a^\dagger_{\pi_2}|{-}>+\sum\limits_{\pi',\nu',\omega}a^\dagger_{\pi'}a^\dagger_{\nu'}A^\dagger_{\omega}(\mu)|{-}>~~~~~~~~~~~~~~~\nonumber\\\times <{-}|[a^\dagger_{\pi'}a^\dagger_{\nu'}A^\dagger_{\omega}(\mu)]^\dagger
			H_{31}a^\dagger_{\pi_1}a^\dagger_{\pi_2}|{-}>E_{\pi_1\pi_2}(\pi'\nu',\omega),
			\label{opn8}
			%	\end{split}
	\end{eqnarray}
	\begin{eqnarray}
		%	\begin{split}
			<{\nu_1\nu_{2corr}}|~=~a^\dagger_{\nu_1}a^\dagger_{\nu_2}|{-}>+\sum\limits_{\pi',\nu',\omega}a^+_{\pi'}a^\dagger_{\nu'}A^\dagger_{\omega}(-\mu)|{-}>~~~~~~~~~~~~~~~\nonumber\\\times <{-}|[a^\dagger_{\pi'}a^\dagger_{\nu'}A^\dagger_{\omega}(-\mu)]^\dagger
			H_{31}a^\dagger_{\nu_1}a^\dagger_{\nu_2}|{-}>E_{\nu_1\nu_2}(\pi'\nu',\omega),
			\label{opn9}
	\end{eqnarray}
	where,
		\begin{equation}
			E_{ab}(cd,\omega)=\frac{1}{(\epsilon_a+\epsilon_b)-(\epsilon_{c}+\epsilon_{d}+\omega)}
			\label{opn10}
		\end{equation}
with subscript index a (b) denotes $\pi,~\pi_1$ and $\nu_1$ ($\nu,~\pi_2$ and $\nu_2$) and c (d) denotes
$\pi',~\pi'_1$ and $\nu'_1$ ($\nu',~\pi'_2$ and $\nu'_2$).
The GT transition amplitudes between these states are reduced to those of one quasi-particle states
	\begin{eqnarray}
	<{\pi_1\pi_{2corr}}|\tau_{\pm}\sigma_{\mu}|{\pi \nu_{corr}}>~=~&\delta(\pi_1,\pi)<{\pi_{2corr}}|\tau_{\pm}\sigma_{\mu}|{\nu_{corr}}>\nonumber\\ &-\delta(\pi_2,\pi)
			<{\pi_{1corr}}|\tau_{\pm}\sigma_{\mu}|{\nu_{corr}}>,
			\label{opn11}
		\end{eqnarray}
	\begin{eqnarray}
			<{\nu_1\nu_{2corr}}|\tau_{\pm}\sigma_{-\mu}|{\pi \nu_{corr}}>~=~&\delta(\nu_2,\nu)<{\nu_{1corr}}|\tau_{\pm}\sigma_{-\mu}|{\pi_{corr}}>\nonumber\\ &-\delta(\nu_1,\nu)
			<{\nu_{2corr}}|\tau_{\pm}\sigma_{-\mu}|{\pi_{corr}}>,
			\label{opn12}
	\end{eqnarray}
by ignoring terms of second order in the correlated phonons.
For odd-odd parent nuclei, QRPA phonon excitations are also possible where the quasi-particle pair acts as spectators in the same single quasi-particle shells. The nuclear excited states can be constructed as phonon correlated multi quasi-particle states. The transition amplitudes between the multi quasi-particle states can be reduced to those of one
quasi-particle states as described below.

Excited levels of an even-even nucleus are two-proton quasi-particle states and two-neutron quasi-particle states.
Transitions from these initial states to final neutron-proton quasi-particle pair states are possible
in the odd-odd daughter nuclei. The transition amplitudes can be reduced to correlated quasi-particle states
by taking the Hermitian conjugate of equations~(\ref{opn11}) and (\ref{opn12})

\begin{eqnarray}
		<{\pi \nu_{corr}}|\tau_{\pm}\sigma_{-\mu}|{\pi_1\pi_{2corr}}>~=~& - \delta(\pi,\pi_2)<{\nu_{corr}}|\tau_{\pm}\sigma_{-\mu}|{\pi_{1corr}}>\nonumber\\&+\delta(\pi,\pi_1)
		<{\nu_{corr}}|\tau_{\pm}\sigma_{-\mu}|{\pi_{2corr}}>,
		\label{opn13}
\end{eqnarray}
\begin{eqnarray}
		<{\pi \nu_{corr}}|\tau_{\pm}\sigma_{\mu}|{\nu_1\nu_{2corr}}>~=~&\delta(\nu,\nu_2)<{\pi_{corr}}|\tau_{\pm}\sigma_{\mu}|{\nu_{1corr}}>\nonumber\\&-\delta(\nu,\nu_1)
		<{\pi_{corr}}|\tau_{\pm}\sigma_{\mu}|{\nu_{2corr}}>.
		\label{opn13}
\end{eqnarray}

When a nucleus has an odd nucleon (a proton and/or a neutron), some low-lying states are obtained by lifting
the quasi-particle in the orbit of the smallest energy to higher-lying orbits. States of an odd-proton even-neutron nucleus
are expressed by three-proton states or one proton two-neutron states, corresponding to excitation
of a proton or a neutron
	\begin{eqnarray}\label{opn14}
		|\pi_1\pi_2\pi_{3corr}\rangle~=~a^\dagger_{\pi_1}a^\dagger_{\pi_2}a^\dagger_{\pi_3}|-\rangle + \frac{1}{2}\sum_{\pi^{'}_1,\pi^{'}_2,\nu^{'},\omega}a^\dagger_{\pi^{'}_1}a^\dagger_{\pi^{'}_2}a^\dagger_{\nu^{'}}A^\dagger_{\omega}(\mu)|-\rangle \nonumber \\
		~~~~~~~~~~~~~~~ \times \langle-|[a^\dagger_{\pi^{'}_1}a^\dagger_{\pi^{'}_2}a^\dagger_{\nu^{'}}A^\dagger_{\omega}(\mu)]^{\dagger}H_{31}a^\dagger_{\pi_1}a^\dagger_{\pi_2}a^\dagger_{\pi_3}|-\rangle \nonumber \\
		~~~~~~~~~~~~~~~ \times E_{\pi_1\pi_2\pi_3}(\pi^{'}_1\pi^{'}_2\nu^{'},\omega)
	\end{eqnarray}
	\begin{eqnarray}\label{opn15}
		|\pi_1\nu_1\nu_{2corr}\rangle ~=~a^\dagger_{\pi_1}a^\dagger_{\nu_1}a^\dagger_{\nu_2}|-\rangle \nonumber + \frac{1}{2}\sum_{\pi^{'}_1,\pi^{'}_2,\nu^{'},\omega}a^\dagger_{\pi^{'}_1}a^\dagger_{\pi^{'}_2}a^\dagger_{\nu^{'}}A^\dagger_{\omega}(-\mu)|-\rangle \nonumber \\
		~~~~~~~~~~~~~~~ \times \langle-|[a^\dagger_{\pi^{'}_1}a^\dagger_{\pi^{'}_2}a^\dagger_{\nu^{'}}A^\dagger_{\omega}(-\mu)]^{\dagger}H_{31}a^\dagger_{\pi_1}a^\dagger_{\nu_1}a^\dagger_{\nu_2}|-\rangle \nonumber \\
		~~~~~~~~~~~~~~~ \times E_{\pi_1\nu_1\nu_2}(\pi^{'}_1\pi^{'}_2\nu^{'},\omega) \nonumber +\frac{1}{6}\sum_{\nu^{'}_1,\nu^{'}_2,\nu^{'}_3,\omega}a^\dagger_{\nu^{'}_1}a^\dagger_{\nu^{'}_2}a^\dagger_{\nu^{'}_3}A^\dagger_{\omega}(\mu)|-\rangle \nonumber \\
		~~~~~~~~~~~~~~~ \times \langle-|[a^\dagger_{\nu^{'}_1}a^\dagger_{\nu^{'}_2}a^\dagger_{\nu^{'}_3}A^\dagger_{\omega}(\mu)]^{\dagger}H_{31}a^\dagger_{\pi_1}a^\dagger_{\nu_1}a^\dagger_{\nu_2}|-\rangle \nonumber \\
		~~~~~~~~~~~~~~~ \times E_{\pi_1\nu_1\nu_2}(\nu^{'}_1\nu^{'}_2\nu^{'}_3,\omega)
	\end{eqnarray}
	with the energy denominators of first order perturbation,
	\begin{equation}
			E_{abc}(def,\omega)=\frac{1}{(\epsilon_{a}+\epsilon_{b}+\epsilon_{c}-\epsilon_{d}-\epsilon_{e}-\epsilon_{f}-\omega)},
			\label{opn16}
		\end{equation}
where subscripts represent $\pi_1$, $\pi_2$, $\pi_3$, $\nu_1$ and $\nu_2$ ($\pi'_1$, $\pi'_2$, $\nu'$, $\nu'_1$, $\nu'_2$ and $\nu'_2$). These equations can be used to generate the three quasi-particle states of odd-proton and even-neutron by swapping the neutron and proton states, $\nu\longleftrightarrow \pi$ and $A^{\dagger}_\omega(\mu) \longleftrightarrow A^{\dagger}_\omega(-\mu)$. Amplitudes of the quasi-particle transitions between the three quasi-particle states are reduced to those for correlated one quasi-particle states. For parent nuclei with an odd proton,
\begin{eqnarray}
		\langle \pi^{'}_1\pi^{'}_2\nu^{'}_{1corr}|\tau_{\pm}\sigma_{-\mu}|\pi_1\pi_2\pi_{3corr}\rangle&
		~=~\delta(\pi^{'}_1,\pi_2)\delta(\pi^{'}_2,\pi_3)\langle \nu^{'}_{1corr}|\tau_{\pm}\sigma_{-\mu}|\pi_{1corr}\rangle \nonumber\\
		~&-\delta(\pi^{'}_1,\pi_1)\delta(\pi^{'}_2,\pi_3)\langle \nu^{'}_{1corr}|\tau_{\pm}\sigma_{-\mu}|\pi_{2corr}\rangle \nonumber\\
		~&+\delta(\pi^{'}_1,\pi_1)\delta(\pi^{'}_2,\pi_2)\langle \nu^{'}_{1corr}|\tau_{\pm}\sigma_{-\mu}|\pi_{3corr}\rangle,
	\end{eqnarray}
	\begin{eqnarray}
		\langle \pi^{'}_1\pi^{'}_2\nu^{'}_{1corr}|\tau_{\pm}\sigma_{\mu}|\pi_1\nu_1\nu_{2corr}\rangle&
		~=~\delta(\nu^{'}_1,\nu_2)[\delta(\pi^{'}_1,\pi_1)\langle \pi^{'}_{2corr}|\tau_{\pm}\sigma_{\mu}|\nu_{1corr}\rangle \nonumber\\
		~&-\delta(\pi^{'}_2,\pi_1)\langle \pi^{'}_{1corr}|\tau_{\pm}\sigma_{\mu}|\nu_{1corr}\rangle] \nonumber\\
		~&-\delta(\nu^{'}_1,\nu_1)[\delta(\pi^{'}_1,\pi_1)\langle \pi^{'}_{2corr}|\tau_{\pm}\sigma_{\mu}|\nu_{2corr}\rangle \nonumber\\
		~&-\delta(\pi^{'}_2,\pi_1)\langle \pi^{'}_{1corr}|\tau_{\pm}\sigma_{\mu}|\nu_{2corr}\rangle],
	\end{eqnarray}
	\begin{eqnarray}
		\langle \nu^{'}_1\nu^{'}_2\nu^{'}_{3corr}|\tau_{\pm}\sigma_{-\mu}|\pi_1\nu_1\nu_{2corr}\rangle &
		~=~\delta(\nu^{'}_2,\nu_1)\delta(\nu^{'}_3,\nu_2)\langle \nu^{'}_{1corr}|\tau_{\pm}\sigma_{-\mu}|\pi_{1corr}\rangle \nonumber\\
		~&-\delta(\nu^{'}_1,\nu_1)\delta(\nu^{'}_3,\nu_2)\langle \nu^{'}_{2corr}|\tau_{\pm}\sigma_{-\mu}|\pi_{1corr}\rangle \nonumber\\
		~&+\delta(\nu^{'}_1,\nu_1)\delta(\nu^{'}_2,\nu_2)\langle \nu^{'}_{3corr}|\tau_{\pm}\sigma_{-\mu}|\pi_{1corr}\rangle,
	\end{eqnarray}
	and for parent nuclei with an odd neutron
	\begin{eqnarray}
		\langle \pi^{'}_1\nu^{'}_1\nu^{'}_{2corr}|\tau_{\pm}\sigma_{\mu}|\nu_1\nu_2\nu_{3corr}\rangle
		&~=~\delta(\nu^{'}_1,\nu_2)\delta(\nu^{'}_2,\nu_3)\langle \pi^{'}_{1corr}|\tau_{\pm}\sigma_{\mu}|\nu_{1corr}\rangle \nonumber\\
		~&-\delta(\nu^{'}_1,\nu_1)\delta(\nu^{'}_2,\nu_3)\langle \pi^{'}_{1corr}|\tau_{\pm}\sigma_{\mu}|\nu_{2corr}\rangle \nonumber\\
		~&+\delta(\nu^{'}_1,\nu_1)\delta(\nu^{'}_2,\nu_2)\langle \pi^{'}_{1corr}|\tau_{\pm}\sigma_{\mu}|\nu_{3corr}\rangle,
	\end{eqnarray}
	\begin{eqnarray}
		\langle \pi^{'}_1\nu^{'}_1\nu^{'}_{2corr}|\tau_{\pm}\sigma_{-\mu}|\pi_1\pi_2\nu_{1corr}\rangle
		&~=~\delta(\pi^{'}_1,\pi_2)[\delta(\nu^{'}_1,\nu_1)\langle \nu^{'}_{2corr}|\tau_{\pm}\sigma_{-\mu}|\pi_{1corr}\rangle \nonumber\\
		&~-\delta(\nu^{'}_2,\nu_1)\langle \nu^{'}_{1corr}|\tau_{\pm}\sigma_{-\mu}|\pi_{1corr}\rangle] \nonumber\\
		~&-\delta(\pi^{'}_1,\pi_1)[\delta(\nu^{'}_1,\nu_1)\langle \nu^{'}_{2corr}|\tau_{\pm}\sigma_{-\mu}|\pi_{2corr}\rangle \nonumber\\
		~&-\delta(\nu^{'}_2,\nu_1)\langle \nu^{'}_{1corr}|\tau_{\pm}\sigma_{-\mu}|\pi_{2corr}\rangle],
	\end{eqnarray}
	\begin{eqnarray}
		\langle \pi^{'}_1\pi^{'}_2\pi^{'}_{3corr}|\tau_{\pm}\sigma_{\mu}|\pi_1\pi_2\nu_{1corr}\rangle
		&~=~\delta(\pi^{'}_2,\pi_1)\delta(\pi^{'}_3,\pi_2)\langle \pi^{'}_{1corr}|\tau_{\pm}\sigma_{\mu}|\nu_{1corr}\rangle \nonumber\\
		&~-\delta(\pi^{'}_1,\pi_1)\delta(\pi^{'}_3,\pi_2)\langle \pi^{'}_{2corr}|\tau_{\pm}\sigma_{\mu}|\nu_{1corr}\rangle \nonumber\\
		~&+\delta(\pi^{'}_1,\pi_1)\delta(\pi^{'}_2,\pi_2)\langle \pi^{'}_{3corr}|\tau_{\pm}\sigma_{\mu}|\nu_{1corr}\rangle.
	\end{eqnarray}
	
Low-lying states in an odd-odd nucleus are expressed
in the quasi-particle picture by proton-neutron pair states (two quasi-particle states) or by states which are
obtained by adding two proton or two-neutron quasi-particles (four quasi-particle states). Transitions from the former
states are described earlier. Phonon-correlated four quasi-particle states can be constructed
similarly to the two and three quasi-particle states. Also in this
case, transition amplitudes for the four quasi-particle states are
reduced into those for the correlated one quasi-particle states	
	\begin{eqnarray}
			<{\pi^{'}_1\pi^{'}_2\nu^{'}_1\nu^{'}_{2corr}}|\tau_{\pm}\sigma_{-\mu}|{\pi_1\pi_2\pi_3\nu_{1corr}}
>~~~~~~~~~~~~~~~~~~~~~~~~~~~~~~~~~~~~~~~~~~~~~~~~~~~~~~~~~~~~~~~~~~~~~~~\nonumber\\~~~~~~~~~~~~~~=~\delta(\nu^{'}_2,\nu_1)[\delta(\pi^{'}_1,\pi_2)\delta(\pi^{'}_2,\pi_3)<{\nu^{'}_{1corr}}|\tau_{\pm}\sigma_{-\mu}|{\pi_{1corr}}>\nonumber\\
			-\delta(\pi^{'}_1,\pi_1)\delta(\pi^{'}_2,\pi_3)<{\nu^{'}_{1corr}}|\tau_{\pm}\sigma_{-\mu}|{\pi_{2corr}}>\nonumber\\+\delta(\pi^{'}_1,\pi_1)\delta(\pi^{'}_2,\pi_2)<{\nu^{'}_{1corr}}|\tau_{\pm}\sigma_{-\mu}|{\pi_{3corr}}>]\nonumber\\
			-\delta(\nu^{'}_1,\nu_1)[\delta(\pi^{'}_1,\pi_2)\delta(\pi^{'}_2,\pi_3)<{\nu^{'}_{2corr}}|\tau_{\pm}\sigma_{-\mu}|{\pi_{1corr}}>\nonumber\\
			-\delta(\pi^{'}_1,\pi_1)\delta(\pi^{'}_2,\pi_3)<{\nu^{'}_{2corr}}|\tau_{\pm}\sigma_{-\mu}|{\pi_{2corr}}>\nonumber\\
			+\delta(\pi^{'}_1,\pi_1)\delta(\pi^{'}_2,\pi_2)<{\nu^{'}_{2corr}}|\tau_{\pm}\sigma_{-\mu}|{\pi_{3corr}}>],
		\end{eqnarray}
	\begin{eqnarray}
			<{\pi^{'}_1\pi^{'}_2\pi^{'}_3\pi^{'}_{4corr}}|\tau_{\pm}\sigma_{\mu}|{\pi_1\pi_2\pi_3\nu_{1corr}}>~~~~~~~~~~~~~~~~~~~~~~~~~~~~~~~~~~~~~~~~~~~~~~~~~~~~~~~~~~~~~~~~~~~~~~~\nonumber\\~~~~~~~~~~~~~~~=~~-\delta(\pi^{'}_2,\pi_1)\delta(\pi^{'}_3,\pi_2)\delta(\pi^{'}_4,\pi_3)<{\pi^{'}_{1corr}}|\tau_{\pm}\sigma_{\mu}|{\nu_{1corr}}>\nonumber\\
			+\delta(\pi^{'}_1,\pi_1)\delta(\pi^{'}_3,\pi_2)\delta(\pi^{'}_4,\pi_3)<{\pi^{'}_{2corr}}|\tau_{\pm}\sigma_{\mu}|{\nu_{1corr}}>\nonumber\\
			-\delta(\pi^{'}_1,\pi_1)\delta(\pi^{'}_2,\pi_2)\delta(\pi^{'}_4,\pi_3)<{\pi^{'}_{3corr}}|\tau_{\pm}\sigma_{\mu}|{\nu_{1corr}}>\nonumber\\
			+\delta(\pi^{'}_1,\pi_1)\delta(\pi^{'}_2,\pi_2)\delta(\pi^{'}_3,\pi_3)<{\pi^{'}_{4corr}}|\tau_{\pm}\sigma_{\mu}|{\nu_{1corr}}>,
				\end{eqnarray}
	\begin{eqnarray}
			<{\pi^{'}_1\pi^{'}_2\nu^{'}_1\nu^{'}_{2corr}}|\tau_{\pm}\sigma_{\mu}|{\pi_1\nu_1\nu_2\nu_{3corr}}>~~~~~~~~~~~~~~~~~~~~~~~~~~~~~~~~~~~~~~~~~~~~~~~~~~~~~~~~~~~~~~~~~~~~~~~\nonumber\\~~~~~~~~~~~~~~~=~~\delta(\pi^{'}_1,\pi_1)[\delta(\nu^{'}_1,\nu_2)\delta(\nu^{'}_2,\nu_3)<{\pi^{'}_{2corr}}|\tau_{\pm}\sigma_{\mu}|{\nu_{1corr}}>\nonumber\\
			-\delta(\nu^{'}_1,\nu_1)\delta(\nu^{'}_2,\nu_3)<{\pi^{'}_{2corr}}|\tau_{\pm}\sigma_{\mu}|{\nu_{2corr}}>\nonumber\\
			+\delta(\nu^{'}_1,\nu_1)\delta(\nu^{'}_2,\nu_2)<{\pi^{'}_{2corr}}|\tau_{\pm}\sigma_{\mu}|{\nu_{3corr}}>]\nonumber\\
			-\delta(\pi^{'}_2,\pi_1)[\delta(\nu^{'}_1,\nu_2)\delta(\nu^{'}_2,\nu_3)<{\pi^{'}_{1corr}}|\tau_{\pm}\sigma_{\mu}|{\nu_{1corr}}>\nonumber\\
			-\delta(\nu^{'}_1,\nu_1)\delta(\nu^{'}_2,\nu_3)<{\pi^{'}_{1corr}}|\tau_{\pm}\sigma_{\mu}|{\nu_{2corr}}>\nonumber\\
			+\delta(\nu^{'}_1,\nu_1)\delta(\nu^{'}_2,\nu_2)<{\pi^{'}_{1corr}}|\tau_{\pm}\sigma_{\mu}|{\nu_{3corr}}>],
	\end{eqnarray}
	\begin{eqnarray}
			<{\nu^{'}_1\nu^{'}_2\nu^{'}_3\nu^{'}_{4corr}}|\tau_{\pm}\sigma_{-\mu}|{\pi_1\nu_1\nu_2\nu_{3corr}}>~~~~~~~~~~~~~~~~~~~~~~~~~~~~~~~~~~~~~~~~~~~~~~~~~~~~~~~~~~~~~~~~~~~~~~~\nonumber\\~~~~~~~~~~~~~~~=~~+\delta(\nu^{'}_2,\nu_1)\delta(\nu^{'}_3,\nu_2)\delta(\nu^{'}_4,\nu_3)<{\nu^{'}_{1corr}}|\tau_{\pm}\sigma_{-\mu}|{\pi_{1corr}}>\nonumber\\
			-\delta(\nu^{'}_1,\nu_1)\delta(\nu^{'}_3,\nu_2)\delta(\nu^{'}_4,\nu_3)<{\nu^{'}_{2corr}}|\tau_{\pm}\sigma_{-\mu}|{\pi_{1corr}}>\nonumber\\
			+\delta(\nu^{'}_1,\nu_1)\delta(\nu^{'}_2,\nu_2)\delta(\nu^{'}_4,\nu_3)<{\nu^{'}_{3corr}}|\tau_{\pm}\sigma_{-\mu}|{\pi_{1corr}}>\nonumber\\
			-\delta(\nu^{'}_1,\nu_1)\delta(\nu^{'}_2,\nu_2)\delta(\nu^{'}_3,\nu_3)<{\nu^{'}_{4corr}}|\tau_{\pm}\sigma_{-\mu}|{\pi_{1corr}}>.
	\end{eqnarray}
The antisymmetrization of the quasi-particles was also taken into account for each of these amplitudes.\\
	$\pi^{'}_4>\pi^{'}_3>\pi^{'}_2>\pi^{'}_1$,  $\nu^{'}_4>\nu^{'}_3>\nu^{'}_2>\nu^{'}_1$,  $\pi_4>\pi_3>\pi_2>\pi_1$,  $\nu_4>\nu_3>\nu_2>\nu_1$.\\
	The GT transitions are taken into account for each phonon's excited state. It is well understood that the quasi-particle in the parent nucleus occupies the same orbit as the excited phonons.  These transitions play an important role in calculating stellar weak rates, which are key inputs for astrophysical simulations.

The form of the Hamiltonian for many-particle QRPA system is:
\begin{equation}
H_{QRPA} = H_{sp} + \hat{V}_{pairing} + \hat{V}_{pp(GT)} + \hat{V}_{ph(GT)},
\label{Ham}
\end{equation}
where $H_{sp}$ is the single-particle Hamiltonian whose energies and wave-vectors are calculated by the deformed Nilsson model~\cite{Nil55}, while $\hat{V}_{pp(GT)}$ (\ref{ppGT}) and $\hat{V}_{ph(GT)}$ (\ref{phGT}) have been introduced earlier in this section. The pairing correlations ($\hat{V}_{pairing}$) are taken into account within BCS formalism with fixed pairing gaps between proton-proton ($\Delta_{\pi\pi}$) and neutron-neutron ($\Delta_{\nu\nu}$) systems. The gaps are determined from the empirical formulae~\cite{Wan12} involving the neutron ($S_{\nu}$) and proton ($S_{\pi}$) separation energies, with the latest experimentally measured energy values~\cite{nds} when available, otherwise, $S_{\pi(\nu)}$ are taken from~\cite{Mol19}. The gap expressions are given by
\begin{eqnarray}
  \Delta_{\pi\pi} &=& \frac{1}{4} (-1)^{1+Z}[S_{\pi}(A+1, Z+1)+S_{\pi}(A-1,Z-1)-2S_{\pi}(A,Z)], \nonumber \\
    \Delta_{\nu\nu} &=& \frac{1}{4} (-1)^{1-Z+A}[S_{\nu}(A+1, Z)+S_{\nu}(A-1,Z)-2S_{\nu}(A,Z)].
\end{eqnarray}

The Nilsson potential parameter~\cite{Nil55}, Nilsson oscillatory constant ($\Omega=\left(45A^{-1/3}-25A^{-2/3}\right)$) and nuclear deformations~\cite{Mol16} are the other required parameters integrated with the proton-neutron QRPA model to perform calculations. In addition, the values of mass excess used here are taken from latest measurements~\cite{nds}.

For any weak transition, total rate is
\begin{eqnarray}
\lambda = \sum _{if}P_{i} \lambda_{if},
\label{Eq:Trate}
\end{eqnarray}
here $i$ denotes parent state and $f$ daughter state which have energies, respectively, E$_{i}$ and E$_{f}$. The summation
is performed over all parent as well as daughter states (up to a maximum of 20 MeV) to achieve required convergence in the values of the calculated rates.
The occupation probability (P$_{i}$) of parent excited states can be computed by using the normal Boltzmann distribution.
The partial rates ($\lambda_{if}$), for any weak transition from $i$ to $f$ state, have dependence upon the phase-space integrals ($\Phi_{if}$) and reduced transition probabilities ($B_{if}$). The expression for $\lambda_{if}$ is
\begin{eqnarray}
\lambda_{if} = \left(\frac{m^{5}_{e}c^{4}g^{2}}{2\hbar^{7} \pi^{3}} \right) \Phi_{if}(E_{fermi}, T, \rho) B_{if}.
\label{Eq:Prate}
\end{eqnarray}
The weak coupling constant ($g$) takes two values; vector ($g_{V}$) and axial-vector ($g_{A}$) coupling constants,
respectively, for axial-vector and vector type transitions. $B_{if}$ has two components, namely; GT ($B(GT_{\pm})_{if}$)
and Fermi ($B(F_{\pm})_{if}$) transition probabilities in BD and EC directions:
\begin{eqnarray}
B_{if} &=& B(F_{\pm})_{if} + \left(\frac{g_{A}}{g_{V}}\right)^{2}B(GT_{\pm})_{if},
\label{Eq:Rprob}
\end{eqnarray}
with $\frac{g_{A}}{g_{V}} = -1.2694$ taken from~\cite{Nak10}, and $B(F)$ and $B(GT)$ are defined by
\begin{eqnarray}
B(GT_{\pm})_{if} &=& \frac{|\langle f||\hat{O}||i \rangle|^{2}}{2J_{i}+1}; \qquad \hat{O} = \sum_{l}\tau^{l}_{\pm}\bsigma^{l},
\label{Eq:FTP}
\end{eqnarray}
and
\begin{eqnarray}
B(F_{\pm})_{if} &=& \frac{|\langle f||\hat{O}||i \rangle|^{2}}{2J_{i}+1}; \qquad \hat{O} = \sum_{l}\tau^{l}_{\pm},
\label{Eq:GTP}
\end{eqnarray}
where $\sigma_{\mu}$ and $\tau_{\pm}$ are, respectively, spin and iso-spin type operators, and other symbols have usual meanings.

The phase-space integrals ($\Phi_{if}$)
in case of BD is
\begin{eqnarray}
\Phi^{BD}_{ij} = \int_{1}^{w_{m}}w(w_{m}-w)^{2}({w^{2} -1)^{1/2}F(+Z,w)(1-G_{-})}dw,
\label{Eq:FIEE}
\end{eqnarray}
while for continuum EC is
\begin{eqnarray}
\Phi^{EC}_{ij} = \int _{w_{l}}^{\infty}w(w_{m}+w)^{2}(w^{2} -1)^{1/2}F(+Z,w)G_{-}dw.
\label{Eq:FIPC}
\end{eqnarray}

In (\ref{Eq:FIEE}) and (\ref{Eq:FIPC}), $w_{m}$ and $w_{l}$ are, respectively, the total BD energy
and EC threshold energy, and $w$ is the total electron energy (rest mass energy + K.E.). The $G_{-}$ represents the Fermi
Dirac distribution functions for electron. The Fermi functions (F) are adopted from~\cite{Gove71}.

\section{Results and Discussions} \label{sec:results}
\begin{figure}
\begin{center}
\includegraphics[width=1.2\textwidth]{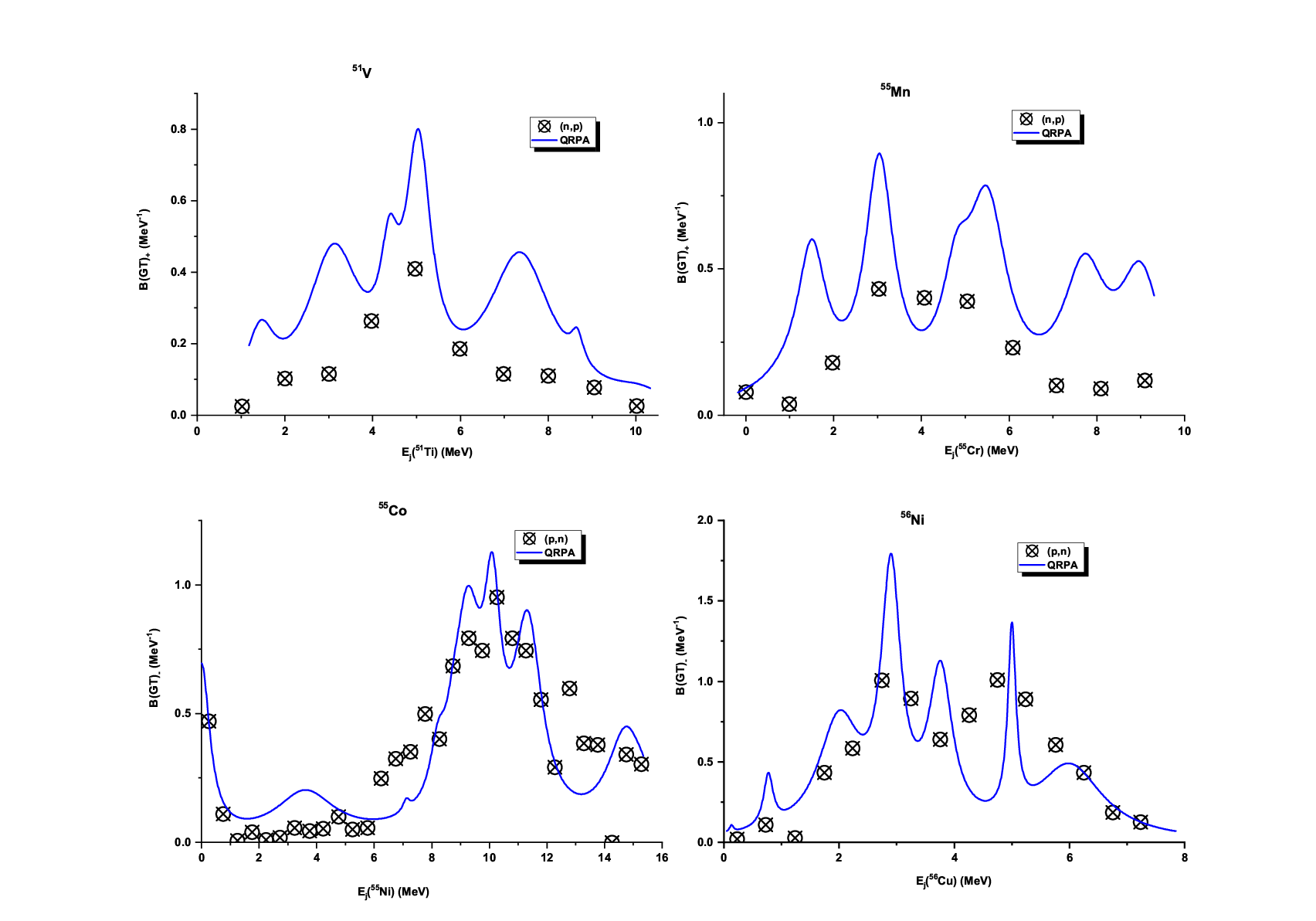}
\end{center}
\caption{\small Comparison of QRPA calculated ground-state
GT strength distributions of $^{51}$V, $^{55}$Mn, $^{55}$Co and $^{56}$Ni with measurements.
For $^{51}$V, experimental data (n,p) was taken from~\cite{Alf93}. The (n,p) data of $^{55}$Mn
was taken from~\cite{Kat94}, whereas  the (p,n) data of $^{55}$Co and $^{56}$Ni was taken from~\cite{Sas12}.
}\label{figure2}
\end{figure}

\begin{figure}
\begin{center}
\includegraphics[width=1.2\textwidth]{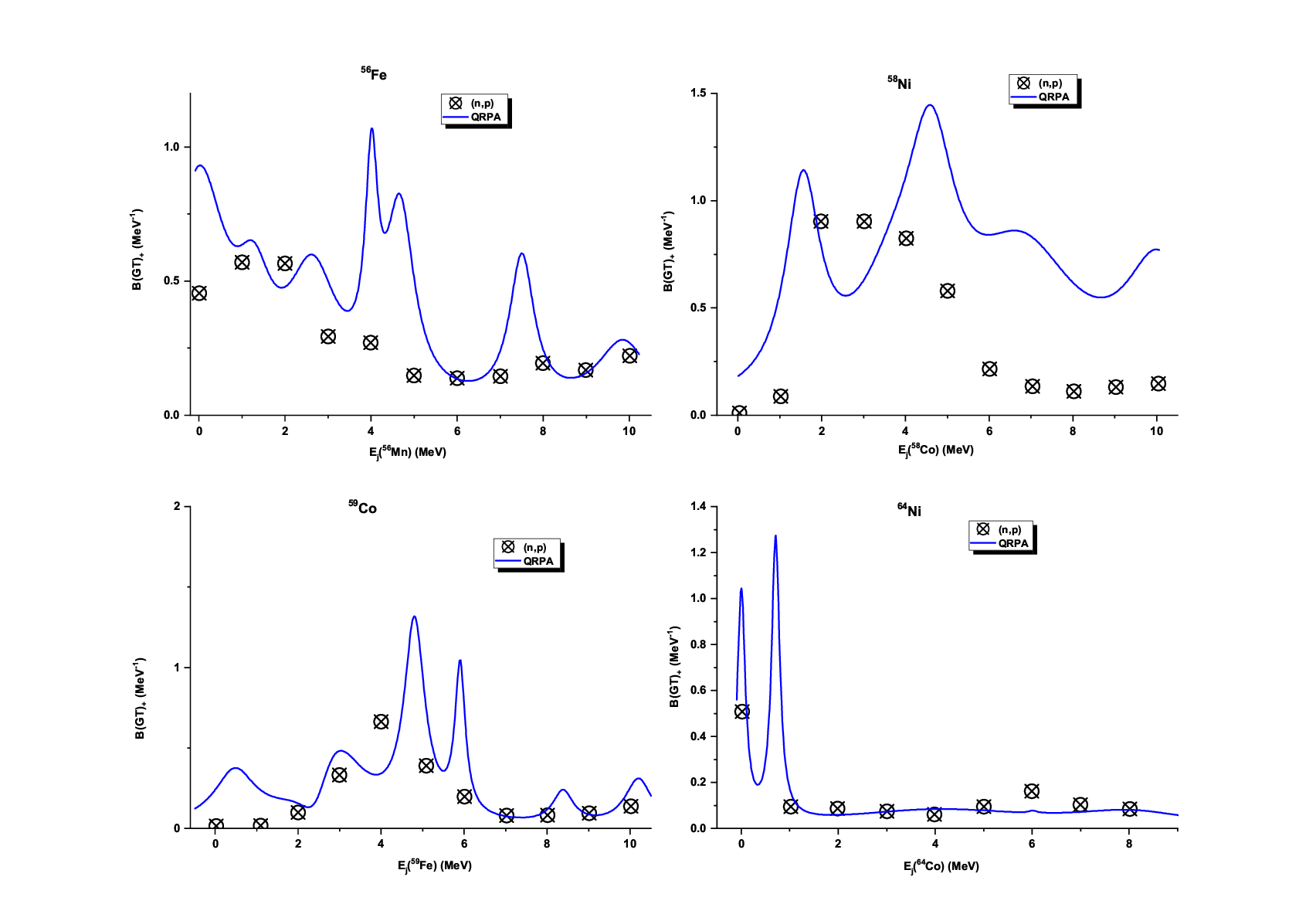}
\end{center}
\caption{\small Same as Figure~\ref{figure2} but for $^{56}$Fe, $^{58}$Ni, $^{59}$Co and $^{64}$Ni.
The measured data (n,p) of $^{56}$Fe and $^{58}$Ni was taken from~\cite{Kat94}, whereas for $^{59}$Co and $^{64}$Ni,
(n,p) data was taken  from~\cite{Alf93} and~\cite{Wil95}, respectively.
}\label{figure3}
\end{figure}

In this section, we present results  on  the effectiveness of the use of the BA hypothesis in
the calculation of GT transitions and weak rates of fp- and fpg-shell nuclei. The selected isotopes were included in the top 50 most relevant nuclei on the basis of their  contribution to the time rate of change of lepton fraction ($\dot{Y_e}$) during presupernova evolution according to a recent simulation study~\cite{Nab21}.  Nuclei with (Z,A) = (20,48) to (34,83) were selected for calculation of weak rates and investigation of BA hypothesis in this project. The cutoff excitation energy in daughter states was 20 MeV.
The comparison of  proton-neutron QRPA calculated ground-state
GT strength distributions of few nuclides with experimental data is shown  in Figures~(\ref{figure2}
and~\ref{figure3}). The theoretical strength distributions have been smeared by a Lorentzian function with an artificial width based on the calculated spectrum.
One should note that the calculated GT strengths are fragmented enough in daughter states.
The calculated strengths of all the presented nuclides show reasonably good comparison with experimental data.
In case of $^{55}$Co, $^{59}$Co and $^{56}$Ni, the calculated GT strengths compare fairly well the measurements.

Our calculations are performed under  astrophysical conditions, relevant to high temperature (kT $>$ 1 MeV ) and density
environments with Fermi energy greater than 10 MeV leading to core-collapse supernovae ( CCSNe). The EC and BD rates under such scenarios
receive significant contribution by many decay channels of low- and high-lying excited states in parent and daughter nuclides. The
total GT centroid and strength values become important~\cite{Fuller} under such circumstances. Many studies regarding such calculations have
often applied the BA hypothesis globally for all the parent excited states above the ground level~\cite{Fuller}, and
other studies have explicitly calculated GT strength distribution for excitation energies up to 1 or 2 MeV in initial levels
and the BA hypothesis is used for the rest of the higher excited levels~\cite{Lan00}. With this consideration, we have put
our selected set of nuclides in three classes. Class 1 contains nuclides with energy values of the first parent
excitation states above 2 MeV calculated by our nuclear model. For class 2 and class 3 nuclides, our model
computes, respectively,
only one initial excited state (above ground level around 1 MeV energy) and several parent excited states with
energies up to 1 MeV. Our class 1 nuclides belong to the global (G)BA hypothesis category where the strength distributions
of \textit{all} initial excited states have been replicated \textit{only} by the ground state GT distribution. Whereas, in class 2
(class 3) nuclides, combined GT strength distributions of the parent ground and first (multiple) excited state(s)
have been replaced for all subsequent levels in the parent nucleus. We refer to class 2 and 3
as localized (L)BA hypothesis.

The main concern of the current study is to show how reliable global and localized BA hypothesis can be in the
theoretical computations of weak decay rates for nuclear astrophysics applications. Therefore, we provide
here a few examples of nuclides for the fulfillment of our aim. We have selected a set of three nuclides containing one
isotope from each class, in the cases of EC and BD transitions, separately.
The selected EC set has $^{78}$Ge, $^{67}$Ni and $^{57}$Ni nuclides,
respectively, from class 1, 2 and 3. Likewise, for BD case, selected set contains $^{58}$Cr, $^{67}$Co and $^{70}$Cu nuclides, in the order from class 1 to 3.
Our adopted methodology to check the reliability of the BA hypothesis generally has three main steps. Firstly,
computations of the total weak rates ($\lambda_{QRPA}$) by using the microscopically calculated GT strength distributions. Secondly, the total weak rates ($\lambda_{G(L)BA}$) have been obtained by implementing
the aforementioned recipe of the GBA (LBA) hypothesis. Total rates in each case have been calculated by using ~(\ref{Eq:Trate}). Lastly, we have computed  Brink-errors ($\Delta_{G(L)BE}$) and standard deviations ($\sigma_{G(L)BE}$)
of the computed rates in order to compare $\lambda_{G(L)BA}$ with the $\lambda_{QRPA}$ and to examine the dependence of total
rates upon microscopically calculated excited states GT strength distribution functions with increasing temperature and density. The algebraic expressions to
calculate $\Delta_{G(L)BE}$ and $\sigma_{G(L)BE}$, are given by (\ref{BE}) and (\ref{SD}), respectively.

\begin{equation}
\Delta_{G(L)BE} = \frac{\lambda_{QRPA} - \lambda_{G(L)BA}}{\lambda_{QRPA}}
\label{BE}
\end{equation}

\begin{equation}
\sigma_{G(L)BE} = \sqrt{\frac{\Sigma {(\Delta_{G(L)BE})}^{2}}{k}}
\label{SD}
\end{equation}
where k represents the total number of temperature-density grid points used in this project.

Tables (\ref{table:table1}$-$\ref{table:table3}) and (\ref{table:table4}$-$\ref{table:table6}) give variations
in $\Delta_{G(L)BE}$ as a function of increasing temperature T (GK) and density $\rho$Y$_{e}$ (g\;cm$^{-3}$), respectively, for EC and BD rates.
In addition, total proton-neutron QRPA
($\lambda_{QRPA}$) and G(L)BA ($\lambda_{G(L)BA}$) weak rates, at each density and temperature, have been presented in
these tables. The rates are given in units of per second.
The values of $\sigma_{G(L)BE}$ in the cases of the GBA and LBA hypothesis have been shown at the top of each table.
Each table shows an increasing trend in $\Delta_{G(L)BE}$ with both increasing temperature and density.
In the cases of EC (table~\ref{table:table1}), $\lambda_{G(L)BA}$ are changed by only several factors
relative to $\lambda_{QRPA}$. Whereas, in the case of BD (table~\ref{table:table4}), weak rates are significantly affected by
incorporating G(L)BA hypothesis. Here the $\lambda_{QRPA}$ rates are modified up to 3 (2) orders of magnitude as compared
to $\lambda_{GBA}$ ($\lambda_{LBA}$). It is noted that at high temperatures, for the case of BD,  the difference between phase spaces based on QRPA and G(L)BA hypothesis formalism is larger than in the case of EC. This contributes to a larger difference between microscopic QRPA and G(L)BA rates for BD nuclei and hence larger magnitude of Brink-errors ($\Delta_{G(L)BE}$) implying poorer BA hypothesis approximation for BD as compared to EC. In both EC and BD cases, $\sigma_{G(L)BE}$ decreases as one goes from GBA rates
to LBA rates. E.g., in BD (EC decay) a standard deviation value with the GBA hypothesis is 330.85 for $^{58}$Cr
(2.71 for $^{78}$Ge). With the LBA hypothesis, the standard deviation value is 24.54 for $^{67}$Co (0.78 for $^{67}$Ni). The use of LBA hypothesis
reduces the deviation between QRPA and BA rates by 1 -- 2 order of magnitude for BD transitions as compared to EC transitions,
where the change between QRPA and BA rates is only a few factors due to the LBA hypothesis.
The cause of increment in the weak rate values is primarily due to the available phase space, lower placement of GT
centroid and/or increase in the total GT strength values.
Table~\ref{table:table7} gives the total GT strength values of weak transitions for EC ($\Sigma B(GT)_{+}$) and BD
($\Sigma B(GT)_{-}$) nuclides, respectively.  Stated also are  the values of GT centroid
($\bar{E}_{\pm}$) for the excited parent states. The negative $\Delta_{G(L)BE}$ values are due to the larger values of
total GT strengths of the ground (and low-lying) state(s) than the total GT strength from the excited (high-lying)
states. These bigger total GT strength values result in the enhancement
of $\lambda_{G(L)BA}$  compared to the $\lambda_{QRPA}$ rates. On the whole, bigger total GT strength values
in the case of BD transitions (compared to EC cases) and differences in available phase space result in a much bigger deviation of $\lambda_{G(L)BA}$
from $\lambda_{QRPA}$ in BD cases and hence, a larger magnitude of $\Delta_{G(L)BE}$ in BD cases than EC ones.

\begin{figure}
\begin{center}
\includegraphics[width=1.2\textwidth]{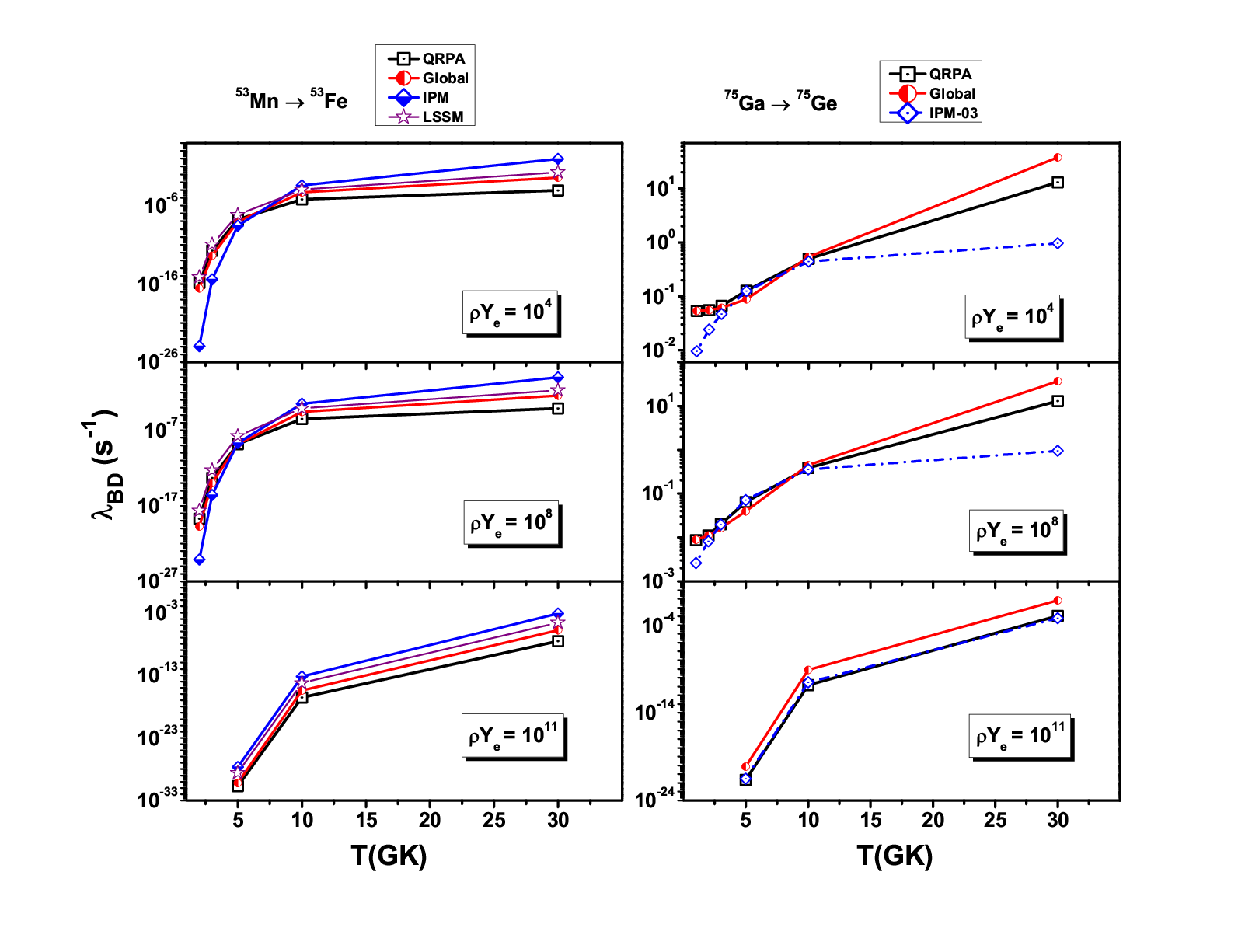}
\end{center}
\caption{\small Comparison of reported BD rates, both QRPA ($\lambda_{QRPA}$) and Global ($\lambda_{GBA}$),
with the previous rates using IPM~\cite{Fuller} and LSSM~\cite{Lan00} (for fp-shell nucleus $^{53}$Mn) and IPM-03~\cite{Pru03}
(for fpg-shell nucleus $^{75}$Ga), at selected astrophysical densities (g\;cm$^{-3}$) and temperatures.
}\label{figure4}
\end{figure}

\begin{figure}
\begin{center}
\includegraphics[width=1.2\textwidth]{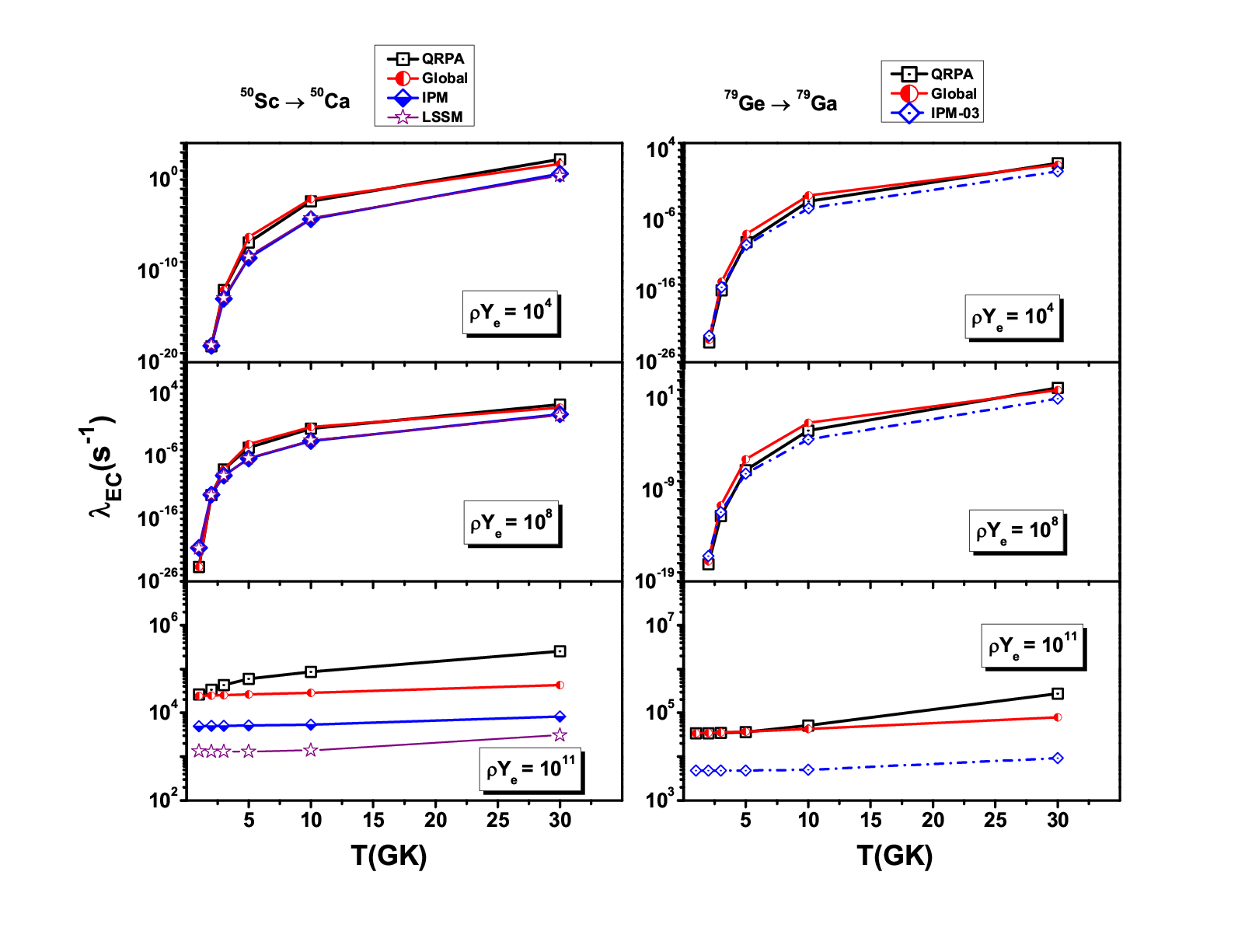}
\end{center}
\caption{\small Same as Figure~\ref{figure4} but for EC rates on fp-shell nucleus $^{50}$Sc and fpg-shell nucleus $^{79}$Ge.
}\label{figure5}
\end{figure}

The previous studies performed by FFN~\cite{Fuller} and Pruet and Fuller~\cite{Pru03} calculated stellar weak rates for
nuclides with A = (21$-$60) and A = (65$-$80), respectively. The BA hypothesis was incorporated in these investigations to estimate the contribution from high-lying states for the rate calculation, using the independent particle model (IPM). Therefore, it is compelling
to compare our QRPA and global rates with these theoretical investigations, to further elaborate the reliability of the BA hypothesis.
Hereafter, in the present paper, the calculations of FFN will be referred to as IPM and that of Pruet and
Fuller as IPM$-$03. We have performed the comparison of our selected fp-
and fpg-shell nuclides, respectively, with IPM and IPM$-$03 in the present study. However, due to
space consideration, the comparison of only one nucleus in each case has been presented, at selected astrophysical
densities [$\rho$Y$_{e}$ = (10$^{4}$, 10$^{8}$ and 10$^{11}$) in units of g\;cm$^{-3}$] and temperatures
[T = (1, 2, 3, 5, 10 and 30) in units of GK]. In addition, we have compared our results of fp-shell nuclei, for the same set of physical conditions, with the earlier reported microscopic rates based on large-scale shell~\cite{Lan00} model (referred to as LSSM). In the LSSM approach, authors explicitly calculated GT
strength distribution for low-lying energy levels and the BA
hypothesis was used for the higher-lying excited levels. LSSM calculations were performed for nuclei in mass range A = (45 - 65). Figure~\ref{figure4} compares our calculated BD rates,
both QRPA ($\lambda_{QRPA}$) and global ($\lambda_{GBA}$), with the corresponding rates of IPM, IPM$-$03 and LSSM.
A similar comparison has been shown in Figure~\ref{figure5} for EC rates.
It is to be noted that in the case of fpg-shell nuclides, our EC (BD)
$\lambda_{QRPA}$ rates are, in general, up to an order of magnitude larger than IPM$-$03 rates,
at all densities and high temperatures. Similar trend are observed between our global $\lambda_{GBA}$ and IPM$-$03 rates.
The BD (EC) rates of IPM and LSSM calculations are bigger (smaller) as compared to both $\lambda_{QRPA}$ and $\lambda_{GBA}$. In the IPM$-$03 approach, GT strengths have been quenched. We did not incorporate any explicit quenching factor in our calculation.
In addition, many states have been included in the partition function sum in the IPM$-$03 calculations, without considering
the corresponding weak transition strengths. The misplacement of GT centroids using the 0$\hbar\omega$ shell model and too small approximated values of unmeasured nuclear matrix elements were few of the weak points adopted in the IPM approach. As  mentioned earlier, LSSM used the BA hypothesis to estimate contribution of rates from higher excited states.
Moreover, LSSM approach applied the Lanczos method to derive GT strength function but was
limited up to 100 iterations which were insufficient for
converging the states above 2.5 MeV excitation energies.
In contrast, the proton neutron QRPA approach computes the GT strengths of all parent states in a
microscopical state-by-state fashion. Due to the availability of a large model space, the pn-QRPA model was able to calculate GT strength distributions from parent excited states  in the computation of stellar rates. The cons of using the current pn-QRPA model are usage of a simple pairing plus quadrupole Hamiltonian  with a schematic and separable interaction. The pn-QRPA model is not based on a thermal phonon vacuum in which the temperature effect is included in the mean-field.
These might be reported as the prime reasons for the differences between QRPA and IPM (IPM$-$03) rates.
Tables (\ref{table:table1}$-$\ref{table:table6}) and Figures~\ref{figure4} and~\ref{figure5} show that, EC and BD rates on
the selected nuclides increase as the core temperature rises. This is because the occupation probability
of the parent excited states rises with increasing temperature and contribute significantly to the total weak rate.  The magnitude of BD (EC)
rates decreases (increases) with rise in density, due to a decrease in available phase space (rise in electron
chemical potential).

We compare our state-by-state EC and BD rates with the LSSM calculation in Table~\ref{table:table8} and Table~\ref{table:table9}, respectively,  at selected densities and temperatures. For the EC case we chose  $^{63}$Ni and for the BD case  $^{63}$Co was selected. Both the selected nuclei belong to Class 3 (LBA).
	
In Table~\ref{table:table8}, we have shown our calculated (This work) EC rates by including ground state (state 1),
lowest 2 (state 2), lowest 3 (state 3) and lowest 5 (state 5) initial states. The corresponding LSSM rates of the lowest 1, 2, 3 and 5 states included microscopically calculated GT distributions. In the last column, we have compared our LBA rates (calculated by incorporating local BA hypothesis) with the LSSM EC rates in which authors included GT distributions calculated from individual states, applied BA hypothesis to derive the remaining GT strength and averaged GT strengths obtained from Lanczos method. On the whole, comparison results of Table~\ref{table:table8} show that state-wise LSSM rates are larger than our calculated rates. However, at high temperatures, the pn-QRPA rates are bigger. The shell model decay rates is a function of the number of Lanczos iterations essential for convergence of rates and this behavior of partition functions can affect their estimates of weak rates at high temperatures. Accordingly at high temperatures, the shell model rates tend to be too low. The Lanczos-based approach used by shell model, is the main cause for this discrepancy and this was also pointed by Ref.~\cite{Pru03}. The pn-QRPA calculations do not suffer from this convergence problem as it is not Lanczos-based approach.

In the LSSM work, authors reported three types of state-by-state rate for BD rates of $^{63}$Co. The rates were calculated using microscopically calculated GT distributions  from
the ground state (state 1) , lowest 2 (state 2) and lowest 4 (state 4) levels of initial nucleus. The comparison between the pn-QRPA and LSSM rates are presented in Table~\ref{table:table9} which shows that the pn-QRPA BD rates surpass the corresponding LSSM rates with increasing excited state GT contributions. The last column shows the BD rates using the two models and invoking BA hypothesis. Once again it is noted that at high temperatures the LSSM rates are too small for reasons discussed before.

\section{Conclusions}
\label{sec:conclusions}
Using the  proton-neutron QRPA model, we investigated how the GT strength functions of allowed transitions and weak rates, in the
EC and BD directions, were affected by the usage of BA hypothesis. Our study covered a broad range of fp- and fpg-shell nuclides with mass numbers (48 -- 83) and proton numbers (20
-- 34). We compared the
microscopically calculated  state-by-state transitions with those obtained by employing the BA hypothesis.
Weak rates, based on the GBA hypothesis, started to deviate from the microscopically calculated
rates at core temperatures exceeding 3 GK and densities beyond $10^{6}$ g$\;$cm$^{-3}$. These physical conditions roughly correlate with  the neon burning phases of stars. A reduction in
the values of standard deviation
was noted by using the LBA hypothesis in place of GBA. The G(L)BA rates were substantially different from the microscopic QRPA rates,  with differences exceeding 3 orders of magnitude. The standard deviation values were factor 100 and beyond bigger in BD transitions than EC cases. At high temperatures, independent of the core density, the difference between phase spaces based on QRPA and G(L)BA hypothesis is larger for BD cases than for EC cases.  This contributes to a larger difference between microscopic QRPA and G(L)BA rates for BD nuclei and hence larger magnitude of Brink-errors.

Based on the present findings, we conclude that the traditionally used BA hypothesis is not an appropriate assumption in obtaining reliable nuclear weak rates at high temperature and density conditions of CCSNe.

\ack J.-U. Nabi would like to acknowledge the support of the Higher Education Commission Pakistan through
Project \# 20-15394/NRPU/R\&D/HEC/2021.

\begin{table}[pt]
\caption{\small  Calculated QRPA ($\lambda_{QRPA}$) and global (G)BA ($\lambda_{GBA}$) EC rates on $^{78}$Ge at different values of stellar temperature, T (GK) and density, $\rho$Y$_{e}$ (g$\;$cm$^{-3}$).  The associated standard deviation ($\sigma_{GBE}$) and Brink-error ($\Delta_{GBE}$) are also given. The rates  are in units of s$^{-1}$.}\label{table:table1}
\setlength{\aboverulesep}{0pt} \setlength{\belowrulesep}{0pt} \setlength{\tabcolsep}{8pt} \setlength{\arrayrulewidth}{5pt}
\setlength\heavyrulewidth{2pt}
{\small
\centering \hspace{20pt}
\begin{tabular}{cccccccc}
\toprule
\multicolumn{8}{|c|}{\rule{0pt}{13pt} \textbf{$^{78}$Ge ($\mathbf{\sigma_{GBE}}$ = 2.71) [Class 1 nucleus]}} \\[0.4ex]
\midrule[2pt]
\midrule[1pt]
\multirow{3}{*}{\textbf{T}}& \multicolumn{3}{c}{\rule{0pt}{13pt} $\mathbf{\brho Y_{e} = 10^{2}}$} &  & \multicolumn{3}{c}{$\mathbf{\brho Y_{e} = 10^{4}}$} \\[0.4ex]
\cmidrule{2-4}  \cmidrule{6-8} &
\multicolumn{1}{c}{$\blambda_{\bf{QRPA}}$} &
\multicolumn{1}{c}{$\blambda_{\bf{GBA}}$} &
\multicolumn{1}{c}{$\bDelta_{\bf{GBE}}$} & &
\multicolumn{1}{c}{$\blambda_{\bf{QRPA}}$} &
\multicolumn{1}{c}{$\blambda_{\bf{GBA}}$} &
\multicolumn{1}{c}{$\bDelta_{\bf{GBE}}$}
\\
\midrule[1pt]
    1     & 5.47$\times$10$^{-49}$ & 1.31$\times$10$^{-48}$ & -1.39 &       & 5.83$\times$10$^{-48}$ & 1.40$\times$10$^{-47}$ & -1.39 \\
    1.5   & 3.54$\times$10$^{-33}$ & 1.50$\times$10$^{-32}$ & -3.24 &       & 5.07$\times$10$^{-33}$ & 2.15$\times$10$^{-32}$ & -3.25 \\
    2     & 4.65$\times$10$^{-25}$ & 2.18$\times$10$^{-24}$ & -3.69 &       & 5.04$\times$10$^{-25}$ & 2.36$\times$10$^{-24}$ & -3.69 \\
    3     & 9.33$\times$10$^{-17}$ & 4.74$\times$10$^{-16}$ & -4.08 &       & 9.46$\times$10$^{-17}$ & 4.81$\times$10$^{-16}$ & -4.08 \\
    5     & 7.74$\times$10$^{-10}$ & 4.41$\times$10$^{-09}$ & -4.69 &       & 7.76$\times$10$^{-10}$ & 4.42$\times$10$^{-09}$ & -4.69 \\
    10    & 3.98$\times$10$^{-04}$ & 2.56$\times$10$^{-03}$ & -5.43 &       & 3.98$\times$10$^{-04}$ & 2.56$\times$10$^{-03}$ & -5.43 \\
    15    & 8.47$\times$10$^{-02}$ & 3.96$\times$10$^{-01}$ & -3.68 &       & 8.47$\times$10$^{-02}$ & 3.96$\times$10$^{-01}$ & -3.68 \\
    20    & 2.24$\times$10$^{+00}$ & 6.24$\times$10$^{+00}$ & -1.78 &       & 2.24$\times$10$^{+00}$ & 6.24$\times$10$^{+00}$ & -1.78 \\
    25    & 2.16$\times$10$^{+01}$ & 3.76$\times$10$^{+01}$ & -0.74 &       & 2.16$\times$10$^{+01}$ & 3.76$\times$10$^{+01}$ & -0.74 \\
    30    & 1.16$\times$10$^{+02}$ & 1.39$\times$10$^{+02}$ & -0.20 &       & 1.16$\times$10$^{+02}$ & 1.39$\times$10$^{+02}$ & -0.20 \\
\midrule[1pt]
\midrule[1pt]
\multirow{3}{*}{\textbf{T}} & \multicolumn{3}{c}{\rule{0pt}{13pt} $\mathbf{\brho Y_{e} = 10^{6}}$} &  & \multicolumn{3}{c}{$\mathbf{\brho Y_{e} = 10^{8}}$} \\[0.4ex]
\cmidrule{2-4}  \cmidrule{6-8} &
\multicolumn{1}{c}{$\blambda_{\bf{QRPA}}$} &
\multicolumn{1}{c}{$\blambda_{\bf{GBA}}$} &
\multicolumn{1}{c}{$\bDelta_{\bf{GBE}}$} & &
\multicolumn{1}{c}{$\blambda_{\bf{QRPA}}$} &
\multicolumn{1}{c}{$\blambda_{\bf{GBA}}$} &
\multicolumn{1}{c}{$\bDelta_{\bf{GBE}}$}
\\
\midrule[1pt]
    1     & 1.26$\times$10$^{-45}$ & 3.02$\times$10$^{-45}$ & -1.39 &       & 9.57$\times$10$^{-37}$ & 2.24$\times$10$^{-36}$ & -1.34 \\
    1.5   & 3.77$\times$10$^{-31}$ & 1.60$\times$10$^{-30}$ & -3.24 &       & 4.89$\times$10$^{-25}$ & 1.98$\times$10$^{-24}$ & -3.06 \\
    2     & 9.08$\times$10$^{-24}$ & 4.26$\times$10$^{-23}$ & -3.69 &       & 5.36$\times$10$^{-19}$ & 2.40$\times$10$^{-18}$ & -3.48 \\
    3     & 2.96$\times$10$^{-16}$ & 1.51$\times$10$^{-15}$ & -4.08 &       & 8.43$\times$10$^{-13}$ & 4.11$\times$10$^{-12}$ & -3.88 \\
    5     & 9.57$\times$10$^{-10}$ & 5.45$\times$10$^{-09}$ & -4.69 &       & 1.25$\times$10$^{-07}$ & 6.92$\times$10$^{-07}$ & -4.52 \\
    10    & 4.07$\times$10$^{-04}$ & 2.62$\times$10$^{-03}$ & -5.43 &       & 2.25$\times$10$^{-03}$ & 1.43$\times$10$^{-02}$ & -5.34 \\
    15    & 8.53$\times$10$^{-02}$ & 3.99$\times$10$^{-01}$ & -3.68 &       & 1.59$\times$10$^{-01}$ & 7.41$\times$10$^{-01}$ & -3.66 \\
    20    & 2.25$\times$10$^{+00}$ & 6.25$\times$10$^{+00}$ & -1.78 &       & 2.95$\times$10$^{+00}$ & 8.17$\times$10$^{+00}$ & -1.77 \\
    25    & 2.17$\times$10$^{+01}$ & 3.77$\times$10$^{+01}$ & -0.74 &       & 2.49$\times$10$^{+01}$ & 4.32$\times$10$^{+01}$ & -0.73 \\
    30    & 1.16$\times$10$^{+02}$ & 1.39$\times$10$^{+02}$ & -0.20 &       & 1.26$\times$10$^{+02}$ & 1.50$\times$10$^{+02}$ & -0.19 \\
\midrule[1pt]
\midrule[1pt]
\multirow{3}{*}{\textbf{T}} & \multicolumn{3}{c}{\rule{0pt}{13pt} $\mathbf{\brho Y_{e} = 10^{10}}$} &  & \multicolumn{3}{c}{$\mathbf{\brho Y_{e} = 10^{11}}$} \\[0.4ex]
\cmidrule{2-4}  \cmidrule{6-8} &
\multicolumn{1}{c}{$\blambda_{\bf{QRPA}}$} &
\multicolumn{1}{c}{$\blambda_{\bf{GBA}}$} &
\multicolumn{1}{c}{$\bDelta_{\bf{GBE}}$} & &
\multicolumn{1}{c}{$\blambda_{\bf{QRPA}}$} &
\multicolumn{1}{c}{$\blambda_{\bf{GBA}}$} &
\multicolumn{1}{c}{$\bDelta_{\bf{GBE}}$}
\\
\midrule[1pt]
    1     & 2.22$\times$10$^{+01}$ & 2.22$\times$10$^{+01}$ & 0.00  &       & 4.70$\times$10$^{+04}$ & 4.70$\times$10$^{+04}$ & 0.00 \\
    1.5   & 2.28$\times$10$^{+01}$ & 2.28$\times$10$^{+01}$ & 0.00  &       & 4.70$\times$10$^{+04}$ & 4.70$\times$10$^{+04}$ & 0.00 \\
    2     & 2.37$\times$10$^{+01}$ & 2.37$\times$10$^{+01}$ & 0.00  &       & 4.70$\times$10$^{+04}$ & 4.70$\times$10$^{+04}$ & 0.00 \\
    3     & 2.62$\times$10$^{+01}$ & 2.62$\times$10$^{+01}$ & 0.00  &       & 4.71$\times$10$^{+04}$ & 4.71$\times$10$^{+04}$ & 0.00 \\
    5     & 3.45$\times$10$^{+01}$ & 3.53$\times$10$^{+01}$ & -0.02 &       & 4.74$\times$10$^{+04}$ & 4.75$\times$10$^{+04}$ & 0.00 \\
    10    & 8.59$\times$10$^{+01}$ & 1.46$\times$10$^{+02}$ & -0.69 &       & 4.97$\times$10$^{+04}$ & 5.26$\times$10$^{+04}$ & -0.06 \\
    15    & 2.57$\times$10$^{+02}$ & 5.46$\times$10$^{+02}$ & -1.12 &       & 6.46$\times$10$^{+04}$ & 6.70$\times$10$^{+04}$ & -0.04 \\
    20    & 7.52$\times$10$^{+02}$ & 1.28$\times$10$^{+03}$ & -0.71 &       & 1.01$\times$10$^{+05}$ & 8.57$\times$10$^{+04}$ & 0.15 \\
    25    & 1.83$\times$10$^{+03}$ & 2.32$\times$10$^{+03}$ & -0.27 &       & 1.54$\times$10$^{+05}$ & 1.05$\times$10$^{+05}$ & 0.32 \\
    30    & 3.75$\times$10$^{+03}$ & 3.63$\times$10$^{+03}$ & 0.03  &       & 2.19$\times$10$^{+05}$ & 1.24$\times$10$^{+05}$ & 0.43 \\
\bottomrule
\end{tabular}}
\end{table}

\begin{table}[pt]
\caption{\small Calculated QRPA ($\lambda_{QRPA}$) and local (L)BA ($\lambda_{LBA}$) EC rates on $^{67}$Ni at different values of stellar temperature, T  (GK) and density, $\rho$Y$_{e}$ (g$\;$cm$^{-3}$).  The associated standard deviation ($\sigma_{LBE}$) and Brink-error ($\Delta_{LBE}$) are also given. The rates  are in units of s$^{-1}$.}\label{table:table2}
\setlength{\aboverulesep}{0pt} \setlength{\belowrulesep}{0pt} \setlength{\tabcolsep}{8pt} \setlength{\arrayrulewidth}{5pt}
\setlength\heavyrulewidth{2pt}
{\small
\centering \hspace{20pt}
\begin{tabular}{cccccccc}
\toprule
\multicolumn{8}{|c|}{\rule{0pt}{13pt} \textbf{$^{67}$Ni ($\mathbf{\sigma_{LBE}}$ = 0.78) [Class 2 nucleus]}} \\[0.4ex]
\midrule[2pt]
\midrule[1pt]
\multirow{3}{*}{\textbf{T}} & \multicolumn{3}{c}{\rule{0pt}{13pt} $\mathbf{\brho Y_{e} = 10^{2}}$} &  & \multicolumn{3}{c}{$\mathbf{\brho Y_{e} = 10^{4}}$} \\[0.4ex]
\cmidrule{2-4}  \cmidrule{6-8} &
\multicolumn{1}{c}{$\blambda_{\bf{QRPA}}$} &
\multicolumn{1}{c}{$\blambda_{\bf{LBA}}$} &
\multicolumn{1}{c}{$\bDelta_{\bf{LBE}}$} & &
\multicolumn{1}{c}{$\blambda_{\bf{QRPA}}$} &
\multicolumn{1}{c}{$\blambda_{\bf{LBA}}$} &
\multicolumn{1}{c}{$\bDelta_{\bf{LBE}}$}
\\
\midrule[1pt]
    1     & 4.74$\times$10$^{-49}$ & 9.25$\times$10$^{-49}$ & -0.95 &       & 5.06$\times$10$^{-48}$ & 9.86$\times$10$^{-48}$ & -0.95 \\
    1.5   & 2.22$\times$10$^{-33}$ & 2.86$\times$10$^{-33}$ & -0.29 &       & 3.18$\times$10$^{-33}$ & 4.10$\times$10$^{-33}$ & -0.29 \\
    2     & 2.87$\times$10$^{-25}$ & 2.13$\times$10$^{-25}$ & 0.26  &       & 3.11$\times$10$^{-25}$ & 2.31$\times$10$^{-25}$ & 0.26 \\
    3     & 6.93$\times$10$^{-17}$ & 2.33$\times$10$^{-17}$ & 0.66  &       & 7.03$\times$10$^{-17}$ & 2.36$\times$10$^{-17}$ & 0.66 \\
    5     & 7.24$\times$10$^{-10}$ & 1.16$\times$10$^{-10}$ & 0.84  &       & 7.26$\times$10$^{-10}$ & 1.16$\times$10$^{-10}$ & 0.84 \\
    10    & 4.05$\times$10$^{-04}$ & 3.72$\times$10$^{-05}$ & 0.91  &       & 4.05$\times$10$^{-04}$ & 3.72$\times$10$^{-05}$ & 0.91 \\
    15    & 6.58$\times$10$^{-02}$ & 4.88$\times$10$^{-03}$ & 0.93  &       & 6.58$\times$10$^{-02}$ & 4.88$\times$10$^{-03}$ & 0.93 \\
    20    & 1.15$\times$10$^{+00}$ & 7.31$\times$10$^{-02}$ & 0.94  &       & 1.15$\times$10$^{+00}$ & 7.31$\times$10$^{-02}$ & 0.94 \\
    25    & 7.46$\times$10$^{+00}$ & 4.23$\times$10$^{-01}$ & 0.94  &       & 7.46$\times$10$^{+00}$ & 4.23$\times$10$^{-01}$ & 0.94 \\
    30    & 2.86$\times$10$^{+01}$ & 1.48$\times$10$^{+00}$ & 0.95  &       & 2.86$\times$10$^{+01}$ & 1.49$\times$10$^{+00}$ & 0.95 \\
\midrule[1pt]
\midrule[1pt]
\multirow{3}{*}{\textbf{T}} & \multicolumn{3}{c}{\rule{0pt}{13pt} $\mathbf{\brho Y_{e} = 10^{6}}$} &  & \multicolumn{3}{c}{$\mathbf{\brho Y_{e} = 10^{8}}$} \\[0.4ex]
\cmidrule{2-4}  \cmidrule{6-8} &
\multicolumn{1}{c}{$\blambda_{\bf{QRPA}}$} &
\multicolumn{1}{c}{$\blambda_{\bf{LBA}}$} &
\multicolumn{1}{c}{$\bDelta_{\bf{LBE}}$} & &
\multicolumn{1}{c}{$\blambda_{\bf{QRPA}}$} &
\multicolumn{1}{c}{$\blambda_{\bf{LBA}}$} &
\multicolumn{1}{c}{$\bDelta_{\bf{LBE}}$}
\\
\midrule[1pt]
    1     & 1.09$\times$10$^{-45}$ & 2.13$\times$10$^{-45}$ & -0.95 &       & 8.55$\times$10$^{-37}$ & 1.67$\times$10$^{-36}$ & -0.95 \\
    1.5   & 2.37$\times$10$^{-31}$ & 3.05$\times$10$^{-31}$ & -0.29 &       & 3.08$\times$10$^{-25}$ & 3.97$\times$10$^{-25}$ & -0.29 \\
    2     & 5.61$\times$10$^{-24}$ & 4.17$\times$10$^{-24}$ & 0.26  &       & 3.32$\times$10$^{-19}$ & 2.46$\times$10$^{-19}$ & 0.26 \\
    3     & 2.20$\times$10$^{-16}$ & 7.40$\times$10$^{-17}$ & 0.66  &       & 6.27$\times$10$^{-13}$ & 2.10$\times$10$^{-13}$ & 0.66 \\
    5     & 8.95$\times$10$^{-10}$ & 1.44$\times$10$^{-10}$ & 0.84  &       & 1.17$\times$10$^{-07}$ & 1.88$\times$10$^{-08}$ & 0.84 \\
    10    & 4.14$\times$10$^{-04}$ & 3.81$\times$10$^{-05}$ & 0.91  &       & 2.29$\times$10$^{-03}$ & 2.10$\times$10$^{-04}$ & 0.91 \\
    15    & 6.62$\times$10$^{-02}$ & 4.91$\times$10$^{-03}$ & 0.93  &       & 1.24$\times$10$^{-01}$ & 9.16$\times$10$^{-03}$ & 0.93 \\
    20    & 1.15$\times$10$^{+00}$ & 7.33$\times$10$^{-02}$ & 0.94  &       & 1.51$\times$10$^{+00}$ & 9.59$\times$10$^{-02}$ & 0.94 \\
    25    & 7.46$\times$10$^{+00}$ & 4.24$\times$10$^{-01}$ & 0.94  &       & 8.57$\times$10$^{+00}$ & 4.86$\times$10$^{-01}$ & 0.94 \\
    30    & 2.86$\times$10$^{+01}$ & 1.49$\times$10$^{+00}$ & 0.95  &       & 3.10$\times$10$^{+01}$ & 1.61$\times$10$^{+00}$ & 0.95 \\

\midrule[1pt]
\midrule[1pt]
\multirow{3}{*}{\textbf{T}} & \multicolumn{3}{c}{\rule{0pt}{13pt} $\mathbf{\brho Y_{e} = 10^{10}}$} &  & \multicolumn{3}{c}{$\mathbf{\brho Y_{e} = 10^{11}}$} \\[0.4ex]
\cmidrule{2-4}  \cmidrule{6-8} &
\multicolumn{1}{c}{$\blambda_{\bf{QRPA}}$} &
\multicolumn{1}{c}{$\blambda_{\bf{LBA}}$} &
\multicolumn{1}{c}{$\bDelta_{\bf{LBE}}$} & &
\multicolumn{1}{c}{$\blambda_{\bf{QRPA}}$} &
\multicolumn{1}{c}{$\blambda_{\bf{LBA}}$} &
\multicolumn{1}{c}{$\bDelta_{\bf{LBE}}$}
\\
\midrule[1pt]
    1     & 1.42$\times$10$^{+00}$ & 1.42$\times$10$^{+00}$ & 0.00  &       & 2.02$\times$10$^{+03}$ & 2.02$\times$10$^{+03}$ & 0.00 \\
    1.5   & 1.36$\times$10$^{+00}$ & 1.34$\times$10$^{+00}$ & 0.02  &       & 2.01$\times$10$^{+03}$ & 2.00$\times$10$^{+03}$ & 0.01 \\
    2     & 1.54$\times$10$^{+00}$ & 1.32$\times$10$^{+00}$ & 0.14  &       & 2.11$\times$10$^{+03}$ & 1.98$\times$10$^{+03}$ & 0.06 \\
    3     & 3.27$\times$10$^{+00}$ & 1.41$\times$10$^{+00}$ & 0.57  &       & 3.03$\times$10$^{+03}$ & 1.98$\times$10$^{+03}$ & 0.35 \\
    5     & 1.32$\times$10$^{+01}$ & 1.92$\times$10$^{+00}$ & 0.86  &       & 7.59$\times$10$^{+03}$ & 2.00$\times$10$^{+03}$ & 0.74 \\
    10    & 6.97$\times$10$^{+01}$ & 5.07$\times$10$^{+00}$ & 0.93  &       & 2.16$\times$10$^{+04}$ & 2.15$\times$10$^{+03}$ & 0.90 \\
    15    & 1.84$\times$10$^{+02}$ & 1.17$\times$10$^{+01}$ & 0.94  &       & 3.32$\times$10$^{+04}$ & 2.32$\times$10$^{+03}$ & 0.93 \\
    20    & 3.69$\times$10$^{+02}$ & 2.16$\times$10$^{+01}$ & 0.94  &       & 4.16$\times$10$^{+04}$ & 2.41$\times$10$^{+03}$ & 0.94 \\
    25    & 6.17$\times$10$^{+02}$ & 3.33$\times$10$^{+01}$ & 0.95  &       & 4.69$\times$10$^{+04}$ & 2.42$\times$10$^{+03}$ & 0.95 \\
    30    & 9.16$\times$10$^{+02}$ & 4.60$\times$10$^{+01}$ & 0.95  &       & 5.02$\times$10$^{+04}$ & 2.40$\times$10$^{+03}$ & 0.95 \\
\bottomrule
\end{tabular}}
\end{table}

\begin{table}[pt]
\caption{\small Same as Table~\ref{table:table2} but for EC rates on $^{57}$Ni.}\label{table:table3}
\setlength{\aboverulesep}{0pt} \setlength{\belowrulesep}{0pt} \setlength{\tabcolsep}{8pt} \setlength{\arrayrulewidth}{5pt}
\setlength\heavyrulewidth{2pt}
{\small
\centering  \hspace{20pt}
\begin{tabular}{cccccccc}
\toprule
\multicolumn{8}{|c|}{\rule{0pt}{13pt} \textbf{$^{57}$Ni ($\mathbf{\sigma_{LBE}}$ = 0.43) [Class 3 nucleus] }} \\[0.4ex]
\midrule[2pt]
\midrule[1pt]
\multirow{3}{*}{\textbf{T}} & \multicolumn{3}{c}{\rule{0pt}{13pt} $\mathbf{\brho Y_{e} = 10^{2}}$} &  & \multicolumn{3}{c}{$\mathbf{\brho Y_{e} = 10^{4}}$} \\[0.4ex]
\cmidrule{2-4}  \cmidrule{6-8} &
\multicolumn{1}{c}{$\blambda_{\bf{QRPA}}$} &
\multicolumn{1}{c}{$\blambda_{\bf{LBA}}$} &
\multicolumn{1}{c}{$\bDelta_{\bf{LBE}}$} & &
\multicolumn{1}{c}{$\blambda_{\bf{QRPA}}$} &
\multicolumn{1}{c}{$\blambda_{\bf{LBA}}$} &
\multicolumn{1}{c}{$\bDelta_{\bf{LBE}}$}
\\
\midrule[1pt]
    1     & 5.79$\times$10$^{-07}$ & 5.79$\times$10$^{-07}$ & 0.00  &       & 6.12$\times$10$^{-06}$ & 6.12$\times$10$^{-06}$ & 0.00 \\
    1.5   & 9.51$\times$10$^{-06}$ & 9.51$\times$10$^{-06}$ & 0.00  &       & 1.36$\times$10$^{-05}$ & 1.36$\times$10$^{-05}$ & 0.00 \\
    2     & 4.88$\times$10$^{-05}$ & 4.88$\times$10$^{-05}$ & 0.00  &       & 5.27$\times$10$^{-05}$ & 5.27$\times$10$^{-05}$ & 0.00 \\
    3     & 3.46$\times$10$^{-04}$ & 3.46$\times$10$^{-04}$ & 0.00  &       & 3.50$\times$10$^{-04}$ & 3.50$\times$10$^{-04}$ & 0.00 \\
    5     & 3.07$\times$10$^{-03}$ & 3.07$\times$10$^{-03}$ & 0.00  &       & 3.07$\times$10$^{-03}$ & 3.08$\times$10$^{-03}$ & 0.00 \\
    10    & 6.50$\times$10$^{-02}$ & 6.44$\times$10$^{-02}$ & 0.01  &       & 6.50$\times$10$^{-02}$ & 6.44$\times$10$^{-02}$ & 0.01 \\
    15    & 7.26$\times$10$^{-01}$ & 5.19$\times$10$^{-01}$ & 0.29  &       & 7.26$\times$10$^{-01}$ & 5.20$\times$10$^{-01}$ & 0.28 \\
    20    & 5.81$\times$10$^{+00}$ & 2.49$\times$10$^{+00}$ & 0.57  &       & 5.81$\times$10$^{+00}$ & 2.49$\times$10$^{+00}$ & 0.57 \\
    25    & 3.03$\times$10$^{+01}$ & 8.15$\times$10$^{+00}$ & 0.73  &       & 3.03$\times$10$^{+01}$ & 8.15$\times$10$^{+00}$ & 0.73 \\
    30    & 1.11$\times$10$^{+02}$ & 2.06$\times$10$^{+01}$ & 0.81  &       & 1.11$\times$10$^{+02}$ & 2.06$\times$10$^{+01}$ & 0.81 \\
\midrule[1pt]
\midrule[1pt]
\multirow{3}{*}{\textbf{T}} & \multicolumn{3}{c}{\rule{0pt}{13pt} $\mathbf{\brho Y_{e} = 10^{6}}$} &  & \multicolumn{3}{c}{$\mathbf{\brho Y_{e} = 10^{8}}$} \\[0.4ex]
\cmidrule{2-4}  \cmidrule{6-8} &
\multicolumn{1}{c}{$\blambda_{\bf{QRPA}}$} &
\multicolumn{1}{c}{$\blambda_{\bf{LBA}}$} &
\multicolumn{1}{c}{$\bDelta_{\bf{LBE}}$} & &
\multicolumn{1}{c}{$\blambda_{\bf{QRPA}}$} &
\multicolumn{1}{c}{$\blambda_{\bf{LBA}}$} &
\multicolumn{1}{c}{$\bDelta_{\bf{LBE}}$}
\\
\midrule[1pt]
    1     & 5.38$\times$10$^{-04}$ & 5.38$\times$10$^{-04}$ & 0.00  &       & 7.64$\times$10$^{-02}$ & 7.64$\times$10$^{-02}$ & 0.00 \\
    1.5   & 6.68$\times$10$^{-04}$ & 6.68$\times$10$^{-04}$ & 0.00  &       & 9.53$\times$10$^{-02}$ & 9.53$\times$10$^{-02}$ & 0.00 \\
    2     & 7.69$\times$10$^{-04}$ & 7.69$\times$10$^{-04}$ & 0.00  &       & 1.10$\times$10$^{-01}$ & 1.10$\times$10$^{-01}$ & 0.00 \\
    3     & 1.03$\times$10$^{-03}$ & 1.03$\times$10$^{-03}$ & 0.00  &       & 1.30$\times$10$^{-01}$ & 1.30$\times$10$^{-01}$ & 0.00 \\
    5     & 3.75$\times$10$^{-03}$ & 3.76$\times$10$^{-03}$ & 0.00  &       & 1.63$\times$10$^{-01}$ & 1.63$\times$10$^{-01}$ & 0.00 \\
    10    & 6.65$\times$10$^{-02}$ & 6.58$\times$10$^{-02}$ & 0.01  &       & 3.18$\times$10$^{-01}$ & 3.14$\times$10$^{-01}$ & 0.01 \\
    15    & 7.31$\times$10$^{-01}$ & 5.22$\times$10$^{-01}$ & 0.29  &       & 1.32$\times$10$^{+00}$ & 9.44$\times$10$^{-01}$ & 0.29 \\
    20    & 5.83$\times$10$^{+00}$ & 2.50$\times$10$^{+00}$ & 0.57  &       & 7.57$\times$10$^{+00}$ & 3.24$\times$10$^{+00}$ & 0.57 \\
    25    & 3.03$\times$10$^{+01}$ & 8.17$\times$10$^{+00}$ & 0.73  &       & 3.47$\times$10$^{+01}$ & 9.31$\times$10$^{+00}$ & 0.73 \\
    30    & 1.11$\times$10$^{+02}$ & 2.06$\times$10$^{+01}$ & 0.81  &       & 1.20$\times$10$^{+02}$ & 2.23$\times$10$^{+01}$ & 0.81 \\
\midrule[1pt]
\midrule[1pt]
\multirow{3}{*}{\textbf{T}} & \multicolumn{3}{c}{\rule{0pt}{13pt} $\mathbf{\brho Y_{e} = 10^{10}}$} &  & \multicolumn{3}{c}{$\mathbf{\brho Y_{e} = 10^{11}}$} \\[0.4ex]
\cmidrule{2-4}  \cmidrule{6-8} &
\multicolumn{1}{c}{$\blambda_{\bf{QRPA}}$} &
\multicolumn{1}{c}{$\blambda_{\bf{LBA}}$} &
\multicolumn{1}{c}{$\bDelta_{\bf{LBE}}$} & &
\multicolumn{1}{c}{$\blambda_{\bf{QRPA}}$} &
\multicolumn{1}{c}{$\blambda_{\bf{LBA}}$} &
\multicolumn{1}{c}{$\bDelta_{\bf{LBE}}$}
\\
\midrule[1pt]
    1     & 9.66$\times$10$^{+01}$ & 9.66$\times$10$^{+01}$ & 0.00  &       & 6.59$\times$10$^{+03}$ & 6.59$\times$10$^{+03}$ & 0.00 \\
    1.5   & 1.06$\times$10$^{+02}$ & 1.06$\times$10$^{+02}$ & 0.00  &       & 6.85$\times$10$^{+03}$ & 6.85$\times$10$^{+03}$ & 0.00 \\
    2     & 1.13$\times$10$^{+02}$ & 1.13$\times$10$^{+02}$ & 0.00  &       & 7.01$\times$10$^{+03}$ & 7.01$\times$10$^{+03}$ & 0.00 \\
    3     & 1.21$\times$10$^{+02}$ & 1.21$\times$10$^{+02}$ & 0.00  &       & 7.16$\times$10$^{+03}$ & 7.16$\times$10$^{+03}$ & 0.00 \\
    5     & 1.30$\times$10$^{+02}$ & 1.30$\times$10$^{+02}$ & 0.00  &       & 7.31$\times$10$^{+03}$ & 7.29$\times$10$^{+03}$ & 0.00 \\
    10    & 1.67$\times$10$^{+02}$ & 1.55$\times$10$^{+02}$ & 0.07  &       & 9.59$\times$10$^{+03}$ & 7.67$\times$10$^{+03}$ & 0.20 \\
    15    & 3.74$\times$10$^{+02}$ & 2.06$\times$10$^{+02}$ & 0.45  &       & 2.46$\times$10$^{+04}$ & 8.51$\times$10$^{+03}$ & 0.65 \\
    20    & 9.16$\times$10$^{+02}$ & 2.83$\times$10$^{+02}$ & 0.69  &       & 5.46$\times$10$^{+04}$ & 9.62$\times$10$^{+03}$ & 0.82 \\
    25    & 1.81$\times$10$^{+03}$ & 3.70$\times$10$^{+02}$ & 0.80  &       & 8.95$\times$10$^{+04}$ & 1.07$\times$10$^{+04}$ & 0.88 \\
    30    & 3.03$\times$10$^{+03}$ & 4.59$\times$10$^{+02}$ & 0.85  &       & 1.23$\times$10$^{+05}$ & 1.16$\times$10$^{+04}$ & 0.91 \\
\bottomrule
\end{tabular}}
\end{table}

\begin{table}[pt]
\caption{\small \centering  Same as Table~\ref{table:table1} but for BD rates on $^{58}$Cr.}\label{table:table4}
\setlength{\aboverulesep}{0pt} \setlength{\belowrulesep}{0pt} \setlength{\tabcolsep}{8pt} \setlength{\arrayrulewidth}{5pt}
\setlength\heavyrulewidth{2pt}
{\small
\centering \hspace{20pt}
\begin{tabular}{cccccccc}
\toprule
\multicolumn{8}{|c|}{\rule{0pt}{13pt} \textbf{$^{58}$Cr ($\mathbf{\sigma_{GBE}}$ = 330.85) [Class 1 nucleus]}} \\[0.4ex]
\midrule[2pt]
\midrule[1pt]
\multirow{3}{*}{\textbf{T}} & \multicolumn{3}{c}{\rule{0pt}{13pt} $\mathbf{\brho Y_{e} = 10^{2}}$} &  & \multicolumn{3}{c}{$\mathbf{\brho Y_{e} = 10^{4}}$} \\[0.4ex]
\cmidrule{2-4}  \cmidrule{6-8} &
\multicolumn{1}{c}{$\blambda_{\bf{QRPA}}$} &
\multicolumn{1}{c}{$\blambda_{\bf{GBA}}$} &
\multicolumn{1}{c}{$\bDelta_{\bf{GBE}}$} & &
\multicolumn{1}{c}{$\blambda_{\bf{QRPA}}$} &
\multicolumn{1}{c}{$\blambda_{\bf{GBA}}$} &
\multicolumn{1}{c}{$\bDelta_{\bf{GBE}}$}
\\
\midrule[1pt]
    1     & 1.30$\times$10$^{-01}$ & 1.30$\times$10$^{-01}$ & 0.00  &       & 1.30$\times$10$^{-01}$ & 1.30$\times$10$^{-01}$ & 0.00 \\
    1.5   & 1.30$\times$10$^{-01}$ & 1.30$\times$10$^{-01}$ & 0.00  &       & 1.30$\times$10$^{-01}$ & 1.30$\times$10$^{-01}$ & 0.00 \\
    2     & 1.30$\times$10$^{-01}$ & 1.30$\times$10$^{-01}$ & 0.00  &       & 1.30$\times$10$^{-01}$ & 1.30$\times$10$^{-01}$ & 0.00 \\
    3     & 1.30$\times$10$^{-01}$ & 1.30$\times$10$^{-01}$ & 0.00  &       & 1.30$\times$10$^{-01}$ & 1.30$\times$10$^{-01}$ & 0.00 \\
    5     & 1.30$\times$10$^{-01}$ & 1.29$\times$10$^{-01}$ & 0.01  &       & 1.30$\times$10$^{-01}$ & 1.29$\times$10$^{-01}$ & 0.01 \\
    10    & 5.21$\times$10$^{-01}$ & 4.56$\times$10$^{-01}$ & 0.13  &       & 5.21$\times$10$^{-01}$ & 4.56$\times$10$^{-01}$ & 0.13 \\
    15    & 2.18$\times$10$^{+00}$ & 4.61$\times$10$^{+01}$ & -20.13 &       & 2.18$\times$10$^{+00}$ & 4.61$\times$10$^{+01}$ & -20.13 \\
    20    & 4.41$\times$10$^{+00}$ & 1.16$\times$10$^{+02}$ & -25.30 &       & 4.41$\times$10$^{+00}$ & 1.16$\times$10$^{+02}$ & -25.30 \\
    25    & 6.50$\times$10$^{+00}$ & 2.31$\times$10$^{+02}$ & -34.56 &       & 6.50$\times$10$^{+00}$ & 2.31$\times$10$^{+02}$ & -34.56 \\
    30    & 8.43$\times$10$^{+00}$ & 3.66$\times$10$^{+02}$ & -42.45 &       & 8.43$\times$10$^{+00}$ & 3.66$\times$10$^{+02}$ & -42.45 \\
\midrule[1pt]
\midrule[1pt]
\multirow{3}{*}{\textbf{T}} & \multicolumn{3}{c}{\rule{0pt}{13pt} $\mathbf{\brho Y_{e} = 10^{6}}$} &  & \multicolumn{3}{c}{$\mathbf{\brho Y_{e} = 10^{8}}$} \\[0.4ex]
\cmidrule{2-4}  \cmidrule{6-8} &
\multicolumn{1}{c}{$\blambda_{\bf{QRPA}}$} &
\multicolumn{1}{c}{$\blambda_{\bf{GBA}}$} &
\multicolumn{1}{c}{$\bDelta_{\bf{GBE}}$} & &
\multicolumn{1}{c}{$\blambda_{\bf{QRPA}}$} &
\multicolumn{1}{c}{$\blambda_{\bf{GBA}}$} &
\multicolumn{1}{c}{$\bDelta_{\bf{GBE}}$}
\\
\midrule[1pt]
    1     & 1.28$\times$10$^{-01}$ & 1.28$\times$10$^{-01}$ & 0.00  &       & 4.35$\times$10$^{-02}$ & 3.45$\times$10$^{-03}$ & 0.92 \\
    1.5   & 1.28$\times$10$^{-01}$ & 1.28$\times$10$^{-01}$ & 0.00  &       & 4.45$\times$10$^{-02}$ & 3.45$\times$10$^{-03}$ & 0.92 \\
    2     & 1.28$\times$10$^{-01}$ & 1.28$\times$10$^{-01}$ & 0.00  &       & 4.58$\times$10$^{-02}$ & 1.15$\times$10$^{-02}$ & 0.75 \\
    3     & 1.28$\times$10$^{-01}$ & 1.28$\times$10$^{-01}$ & 0.00  &       & 4.94$\times$10$^{-02}$ & 1.48$\times$10$^{-02}$ & 0.70 \\
    5     & 1.29$\times$10$^{-01}$ & 1.29$\times$10$^{-01}$ & 0.01  &       & 6.18$\times$10$^{-02}$ & 1.53$\times$10$^{-02}$ & 0.75 \\
    10    & 5.21$\times$10$^{-01}$ & 4.55$\times$10$^{-01}$ & 0.13  &       & 4.48$\times$10$^{-01}$ & 2.48$\times$10$^{-02}$ & 0.94 \\
    15    & 2.18$\times$10$^{+00}$ & 4.61$\times$10$^{+01}$ & -20.13 &       & 2.06$\times$10$^{+00}$ & 4.57$\times$10$^{+01}$ & -21.18 \\
    20    & 4.41$\times$10$^{+00}$ & 1.16$\times$10$^{+02}$ & -25.30 &       & 4.26$\times$10$^{+00}$ & 1.14$\times$10$^{+02}$ & -25.85 \\
    25    & 6.50$\times$10$^{+00}$ & 2.31$\times$10$^{+02}$ & -34.56 &       & 6.37$\times$10$^{+00}$ & 2.29$\times$10$^{+02}$ & -34.97 \\
    30    & 8.43$\times$10$^{+00}$ & 3.66$\times$10$^{+02}$ & -42.45 &       & 8.32$\times$10$^{+00}$ & 3.63$\times$10$^{+02}$ & -42.65 \\
\midrule[1pt]
\midrule[1pt]
\multirow{3}{*}{\textbf{T}} & \multicolumn{3}{c}{\rule{0pt}{13pt} $\mathbf{\brho Y_{e} = 10^{10}}$} &  & \multicolumn{3}{c}{$\mathbf{\brho Y_{e} = 10^{11}}$} \\[0.4ex]
\cmidrule{2-4}  \cmidrule{6-8} &
\multicolumn{1}{c}{$\blambda_{\bf{QRPA}}$} &
\multicolumn{1}{c}{$\blambda_{\bf{GBA}}$} &
\multicolumn{1}{c}{$\bDelta_{\bf{GBE}}$} & &
\multicolumn{1}{c}{$\blambda_{\bf{QRPA}}$} &
\multicolumn{1}{c}{$\blambda_{\bf{GBA}}$} &
\multicolumn{1}{c}{$\bDelta_{\bf{GBE}}$}
\\
\midrule[1pt]
    1     & 5.47$\times$10$^{-38}$ & 5.26$\times$10$^{-38}$ & 0.04  &       & 1.00$\times$10$^{-100}$ & 1.00$\times$10$^{-100}$ & 0.00 \\
    1.5   & 4.68$\times$10$^{-26}$ & 7.41$\times$10$^{-26}$ & -0.58 &       & 4.44$\times$10$^{-69}$ & 7.96$\times$10$^{-67}$ & -178.47 \\
    2     & 5.73$\times$10$^{-20}$ & 1.16$\times$10$^{-19}$ & -1.03 &       & 3.19$\times$10$^{-52}$ & 7.14$\times$10$^{-50}$ & -222.87 \\
    3     & 1.06$\times$10$^{-13}$ & 2.72$\times$10$^{-13}$ & -1.57 &       & 3.44$\times$10$^{-35}$ & 9.33$\times$10$^{-33}$ & -270.02 \\
    5     & 2.23$\times$10$^{-08}$ & 6.52$\times$10$^{-08}$ & -1.92 &       & 2.88$\times$10$^{-21}$ & 1.06$\times$10$^{-18}$ & -367.13 \\
    10    & 6.78$\times$10$^{-04}$ & 4.48$\times$10$^{-02}$ & -65.07 &       & 2.32$\times$10$^{-10}$ & 1.12$\times$10$^{-07}$ & -480.95 \\
    15    & 3.08$\times$10$^{-02}$ & 2.38$\times$10$^{+00}$ & -76.45 &       & 1.44$\times$10$^{-06}$ & 1.04$\times$10$^{-03}$ & -726.78 \\
    20    & 2.17$\times$10$^{-01}$ & 2.46$\times$10$^{+01}$ & -112.50 &       & 1.18$\times$10$^{-04}$ & 1.11$\times$10$^{-01}$ & -934.41 \\
    25    & 7.11$\times$10$^{-01}$ & 9.79$\times$10$^{+01}$ & -136.72 &       & 1.71$\times$10$^{-03}$ & 1.94$\times$10$^{+00}$ & -1136.63 \\
    30    & 1.60$\times$10$^{+00}$ & 1.00$\times$10$^{+02}$ & -61.52 &       & 1.05$\times$10$^{-02}$ & 1.90$\times$10$^{+01}$ & -1814.52 \\
\bottomrule
\end{tabular}}
\end{table}

\begin{table}[pt]
\caption{\small \centering Same as Table~\ref{table:table2} but for BD rates on $^{67}$Co.}\label{table:table5}
\setlength{\aboverulesep}{0pt} \setlength{\belowrulesep}{0pt} \setlength{\tabcolsep}{8pt} \setlength{\arrayrulewidth}{5pt}
\setlength\heavyrulewidth{2pt}
{\small
\centering \hspace{20pt}
\begin{tabular}{cccccccc}
\toprule
\multicolumn{8}{|c|}{\rule{0pt}{13pt} \textbf{$^{67}$Co ($\mathbf{\sigma_{LBE}}$ = 24.54) [Class 2 nucleus] }} \\[0.4ex]
\midrule[2pt]
\midrule[1pt]
\multirow{3}{*}{\textbf{T}} & \multicolumn{3}{c}{\rule{0pt}{13pt} $\mathbf{\brho Y_{e} = 10^{2}}$} &  & \multicolumn{3}{c}{$\mathbf{\brho Y_{e} = 10^{4}}$} \\[0.4ex]
\cmidrule{2-4}  \cmidrule{6-8} &
\multicolumn{1}{c}{$\blambda_{\bf{QRPA}}$} &
\multicolumn{1}{c}{$\blambda_{\bf{LBA}}$} &
\multicolumn{1}{c}{$\bDelta_{\bf{LBE}}$} & &
\multicolumn{1}{c}{$\blambda_{\bf{QRPA}}$} &
\multicolumn{1}{c}{$\blambda_{\bf{LBA}}$} &
\multicolumn{1}{c}{$\bDelta_{\bf{LBE}}$}
\\
\midrule[1pt]
    1     & 5.82$\times$10$^{+00}$ & 5.82$\times$10$^{+00}$ & 0.00  &       & 5.82$\times$10$^{+00}$ & 5.82$\times$10$^{+00}$ & 0.00 \\
    1.5   & 6.17$\times$10$^{+00}$ & 6.17$\times$10$^{+00}$ & 0.00  &       & 6.17$\times$10$^{+00}$ & 6.17$\times$10$^{+00}$ & 0.00 \\
    2     & 6.53$\times$10$^{+00}$ & 6.53$\times$10$^{+00}$ & 0.00  &       & 6.53$\times$10$^{+00}$ & 6.53$\times$10$^{+00}$ & 0.00 \\
    3     & 7.18$\times$10$^{+00}$ & 7.18$\times$10$^{+00}$ & 0.00  &       & 7.18$\times$10$^{+00}$ & 7.18$\times$10$^{+00}$ & 0.00 \\
    5     & 8.00$\times$10$^{+00}$ & 8.07$\times$10$^{+00}$ & -0.01 &       & 8.00$\times$10$^{+00}$ & 8.07$\times$10$^{+00}$ & -0.01 \\
    10    & 1.28$\times$10$^{+01}$ & 1.24$\times$10$^{+01}$ & 0.03  &       & 1.28$\times$10$^{+01}$ & 1.24$\times$10$^{+01}$ & 0.03 \\
    15    & 2.75$\times$10$^{+01}$ & 2.56$\times$10$^{+01}$ & 0.07  &       & 2.75$\times$10$^{+01}$ & 2.56$\times$10$^{+01}$ & 0.07 \\
    20    & 4.86$\times$10$^{+01}$ & 4.98$\times$10$^{+01}$ & -0.02 &       & 4.86$\times$10$^{+01}$ & 4.98$\times$10$^{+01}$ & -0.02 \\
    25    & 7.26$\times$10$^{+01}$ & 8.39$\times$10$^{+01}$ & -0.16 &       & 7.26$\times$10$^{+01}$ & 8.39$\times$10$^{+01}$ & -0.16 \\
    30    & 9.71$\times$10$^{+01}$ & 1.25$\times$10$^{+02}$ & -0.29 &       & 9.71$\times$10$^{+01}$ & 1.25$\times$10$^{+02}$ & -0.29 \\
\midrule[1pt]
\midrule[1pt]
\multirow{3}{*}{\textbf{T}} & \multicolumn{3}{c}{\rule{0pt}{13pt} $\mathbf{\brho Y_{e} = 10^{6}}$} &  & \multicolumn{3}{c}{$\mathbf{\brho Y_{e} = 10^{8}}$} \\[0.4ex]
\cmidrule{2-4}  \cmidrule{6-8} &
\multicolumn{1}{c}{$\blambda_{\bf{QRPA}}$} &
\multicolumn{1}{c}{$\blambda_{\bf{LBA}}$} &
\multicolumn{1}{c}{$\bDelta_{\bf{LBE}}$} & &
\multicolumn{1}{c}{$\blambda_{\bf{QRPA}}$} &
\multicolumn{1}{c}{$\blambda_{\bf{LBA}}$} &
\multicolumn{1}{c}{$\bDelta_{\bf{LBE}}$}
\\
\midrule[1pt]
    1     & 5.79$\times$10$^{+00}$ & 5.79$\times$10$^{+00}$ & 0.00  &       & 4.82$\times$10$^{+00}$& 4.82$\times$10$^{+00}$ & 0.00 \\
    1.5   & 6.14$\times$10$^{+00}$ & 6.14$\times$10$^{+00}$ & 0.00  &       & 5.12$\times$10$^{+00}$ & 5.12$\times$10$^{+00}$ & 0.00 \\
    2     & 6.52$\times$10$^{+00}$ & 6.52$\times$10$^{+00}$ & 0.00  &       & 5.43$\times$10$^{+00}$ & 5.43$\times$10$^{+00}$ & 0.00 \\
    3     & 7.16$\times$10$^{+00}$ & 7.16$\times$10$^{+00}$ & 0.00  &       & 6.01$\times$10$^{+00}$ & 6.01$\times$10$^{+00}$ & 0.00 \\
    5     & 8.00$\times$10$^{+00}$ & 8.07$\times$10$^{+00}$ & -0.01 &       & 6.84$\times$10$^{+00}$ & 6.90$\times$10$^{+00}$ & -0.01 \\
    10    & 1.28$\times$10$^{+01}$ & 1.24$\times$10$^{+01}$ & 0.03  &       & 1.17$\times$10$^{+01}$ & 1.14$\times$10$^{+01}$ & 0.02 \\
    15    & 2.75$\times$10$^{+01}$ & 2.56$\times$10$^{+01}$ & 0.07  &       & 2.61$\times$10$^{+01}$ & 2.47$\times$10$^{+01}$ & 0.05 \\
    20    & 4.86$\times$10$^{+01}$ & 4.98$\times$10$^{+01}$ & -0.02 &       & 4.72$\times$10$^{+01}$ & 4.89$\times$10$^{+01}$ & -0.04 \\
    25    & 7.26$\times$10$^{+01}$ & 8.39$\times$10$^{+01}$ & -0.16 &       & 7.13$\times$10$^{+01}$ & 8.30$\times$10$^{+01}$ & -0.16 \\
    30    & 9.71$\times$10$^{+01}$ & 1.25$\times$10$^{+02}$ & -0.29 &       & 9.57$\times$10$^{+01}$ & 1.24$\times$10$^{+02}$ & -0.29 \\
\midrule[1pt]
\midrule[1pt]
\multirow{3}{*}{\textbf{T}} & \multicolumn{3}{c}{\rule{0pt}{13pt} $\mathbf{\brho Y_{e} = 10^{10}}$} &  & \multicolumn{3}{c}{$\mathbf{\brho Y_{e} = 10^{11}}$} \\[0.4ex]
\cmidrule{2-4}  \cmidrule{6-8} &
\multicolumn{1}{c}{$\blambda_{\bf{QRPA}}$} &
\multicolumn{1}{c}{$\blambda_{\bf{LBA}}$} &
\multicolumn{1}{c}{$\bDelta_{\bf{LBE}}$} & &
\multicolumn{1}{c}{$\blambda_{\bf{QRPA}}$} &
\multicolumn{1}{c}{$\blambda_{\bf{LBA}}$} &
\multicolumn{1}{c}{$\bDelta_{\bf{LBE}}$}
\\
\midrule[1pt]
    1     & 3.10$\times$10$^{-15}$ & 6.01$\times$10$^{-15}$ & -0.94 &       & 8.49$\times$10$^{-80}$ & 6.46$\times$10$^{-78}$ & -75.03 \\
    1.5   & 4.94$\times$10$^{-11}$ & 9.79$\times$10$^{-11}$ & -0.98 &       & 4.60$\times$10$^{-54}$ & 3.52$\times$10$^{-52}$ & -75.56 \\
    2     & 8.13$\times$10$^{-09}$ & 1.71$\times$10$^{-08}$& -1.11 &       & 4.38$\times$10$^{-41}$ & 3.41$\times$10$^{-39 }$& -76.98 \\
    3     & 1.91$\times$10$^{-06}$ & 4.93$\times$10$^{-06}$ & -1.58 &       & 5.87$\times$10$^{-28}$ & 4.79$\times$10$^{-26}$ & -80.47 \\
    5     & 2.62$\times$10$^{-04}$ & 1.04$\times$10$^{-03}$ & -2.98 &       & 3.18$\times$10$^{-17}$ & 2.77$\times$10$^{-15}$ & -86.10 \\
    10    & 3.62$\times$10$^{-02}$ & 2.04$\times$10$^{-01}$ & -4.62 &       & 1.20$\times$10$^{-08}$ & 7.50$\times$10$^{-07}$ & -61.37 \\
    15    & 5.11$\times$10$^{-01}$ & 2.18$\times$10$^{+00}$ & -3.28 &       & 2.33$\times$10$^{-05}$ & 6.81$\times$10$^{-04}$ & -28.17 \\
    20    & 2.69$\times$10$^{+00}$ & 9.16$\times$10$^{+00}$ & -2.41 &       & 1.43$\times$10$^{-03}$ & 2.30$\times$10$^{-02}$ & -15.14 \\
    25    & 8.24$\times$10$^{+00}$ & 2.41$\times$10$^{+01}$ & -1.92 &       & 1.92$\times$10$^{-02}$ & 2.01$\times$10$^{-01}$ & -9.50 \\
    30    & 1.84$\times$10$^{+01}$ & 4.80$\times$10$^{+01}$ & -1.61 &       & 1.16$\times$10$^{-01}$ & 8.87$\times$10$^{-01}$ & -6.67 \\
\bottomrule
\end{tabular}}
\end{table}

\begin{table}[pt]
\caption{\small \centering Same as Table~\ref{table:table2} but for BD rates on $^{70}$Cu.}\label{table:table6}
\setlength{\aboverulesep}{0pt} \setlength{\belowrulesep}{0pt} \setlength{\tabcolsep}{8pt} \setlength{\arrayrulewidth}{5pt}
\setlength\heavyrulewidth{2pt}
{\small
\centering \hspace{20pt}
\begin{tabular}{cccccccc}
\toprule
\multicolumn{8}{|c|}{\rule{0pt}{13pt} \textbf{$^{70}$Cu ($\mathbf{\sigma_{LBE}}$ = 13.58) [Class 3 nucleus]} } \\[0.4ex]
\midrule[2pt]
\midrule[1pt]
\multirow{3}{*}{\textbf{T}} & \multicolumn{3}{c}{\rule{0pt}{13pt} $\mathbf{\brho Y_{e} = 10^{2}}$} &  & \multicolumn{3}{c}{$\mathbf{\brho Y_{e} = 10^{4}}$} \\[0.4ex]
\cmidrule{2-4}  \cmidrule{6-8} &
\multicolumn{1}{c}{$\blambda_{\bf{QRPA}}$} &
\multicolumn{1}{c}{$\blambda_{\bf{LBA}}$} &
\multicolumn{1}{c}{$\bDelta_{\bf{LBE}}$} & &
\multicolumn{1}{c}{$\blambda_{\bf{QRPA}}$} &
\multicolumn{1}{c}{$\blambda_{\bf{LBA}}$} &
\multicolumn{1}{c}{$\bDelta_{\bf{LBE}}$}
\\
\midrule[1pt]
    1     & 4.82$\times$10$^{-01}$ & 4.82$\times$10$^{-01}$ & 0.00  &       & 4.82$\times$10$^{-01}$ & 4.82$\times$10$^{-01}$ & 0.00 \\
    1.5   & 6.25$\times$10$^{-01}$ & 6.25$\times$10$^{-01}$ & 0.00  &       & 6.25$\times$10$^{-01}$ & 6.25$\times$10$^{-01}$ & 0.00 \\
    2     & 7.50$\times$10$^{-01}$ & 7.50$\times$10$^{-01}$ & 0.00  &       & 7.50$\times$10$^{-01}$ & 7.50$\times$10$^{-01}$ & 0.00 \\
    3     & 9.53$\times$10$^{-01}$ & 9.55$\times$10$^{-01}$ & 0.00  &       & 9.53$\times$10$^{-01}$ & 9.55$\times$10$^{-01}$ & 0.00 \\
    5     & 1.17$\times$10$^{+00}$ & 1.24$\times$10$^{+00}$ & -0.06 &       & 1.17$\times$10$^{+00}$ & 1.24$\times$10$^{+00}$ & -0.06 \\
    10    & 1.35$\times$10$^{+00}$ & 2.56$\times$10$^{+00}$ & -0.91 &       & 1.35$\times$10$^{+00}$ & 2.56$\times$10$^{+00}$ & -0.91 \\
    15    & 1.91$\times$10$^{+00}$ & 6.75$\times$10$^{+00}$ & -2.54 &       & 1.91$\times$10$^{+00}$ & 6.75$\times$10$^{+00}$ & -2.54 \\
    20    & 2.89$\times$10$^{+00}$ & 1.40$\times$10$^{+01}$ & -3.85 &       & 2.89$\times$10$^{+00}$& 1.40$\times$10$^{+01}$ & -3.85 \\
    25    & 4.02$\times$10$^{+00}$ & 2.27$\times$10$^{+01}$ & -4.65 &       & 4.02$\times$10$^{+00}$ & 2.27$\times$10$^{+01}$ & -4.65 \\
    30    & 5.09$\times$10$^{+00}$ & 3.12$\times$10$^{+01}$ & -5.12 &       & 5.09$\times$10$^{+00}$ & 3.12$\times$10$^{+01}$ & -5.12 \\
\midrule[1pt]
\midrule[1pt]
\multirow{3}{*}{\textbf{T}} & \multicolumn{3}{c}{\rule{0pt}{13pt} $\mathbf{\brho Y_{e} = 10^{6}}$} &  & \multicolumn{3}{c}{$\mathbf{\brho Y_{e} = 10^{8}}$} \\[0.4ex]
\cmidrule{2-4}  \cmidrule{6-8} &
\multicolumn{1}{c}{$\blambda_{\bf{QRPA}}$} &
\multicolumn{1}{c}{$\blambda_{\bf{LBA}}$} &
\multicolumn{1}{c}{$\bDelta_{\bf{LBE}}$} & &
\multicolumn{1}{c}{$\blambda_{\bf{QRPA}}$} &
\multicolumn{1}{c}{$\blambda_{\bf{LBA}}$} &
\multicolumn{1}{c}{$\bDelta_{\bf{LBE}}$}
\\
\midrule[1pt]
    1     & 4.72$\times$10$^{-01}$ & 4.72$\times$10$^{-01}$ & 0.00  &       & 1.68$\times$10$^{-01}$ & 1.68$\times$10$^{-01}$ & 0.00 \\
    1.5   & 6.14$\times$10$^{-01}$ & 6.14$\times$10$^{-01}$ & 0.00  &       & 2.26$\times$10$^{-01}$ & 2.26$\times$10$^{-01}$ & 0.00 \\
    2     & 7.38$\times$10$^{-01}$ & 7.38$\times$10$^{-01}$ & 0.00  &       & 2.83$\times$10$^{-01}$ & 2.83$\times$10$^{-01}$ & 0.00 \\
    3     & 9.42$\times$10$^{-01}$ & 9.44$\times$10$^{-01}$ & 0.00  &       & 3.94$\times$10$^{-01}$ & 3.95$\times$10$^{-01}$ & 0.00 \\
    5     & 1.16$\times$10$^{+00}$ & 1.23$\times$10$^{+00}$ & -0.06 &       & 5.83$\times$10$^{-01}$ & 6.35$\times$10$^{-01}$ & -0.09 \\
    10    & 1.34$\times$10$^{+00}$ & 2.56$\times$10$^{+00}$ & -0.91 &       & 9.91$\times$10$^{-01}$ & 2.07$\times$10$^{+00}$ & -1.09 \\
    15    & 1.91$\times$10$^{+00}$ & 6.75$\times$10$^{+00}$ & -2.54 &       & 1.67$\times$10$^{+00}$ & 6.28$\times$10$^{+00}$ & -2.75 \\
    20    & 2.88$\times$10$^{+00}$ & 1.40$\times$10$^{+01}$ & -3.86 &       & 2.71$\times$10$^{+00}$ & 1.36$\times$10$^{+01}$ & -4.00 \\
    25    & 4.02$\times$10$^{+00}$ & 2.27$\times$10$^{+01}$ & -4.65 &       & 3.86$\times$10$^{+00}$ & 2.22$\times$10$^{+01}$ & -4.75 \\
    30    & 5.09$\times$10$^{+00}$ & 3.12$\times$10$^{+01}$ & -5.12 &       & 4.98$\times$10$^{+00}$ & 3.08$\times$10$^{+01}$ & -5.18 \\
\midrule[1pt]
\midrule[1pt]
\multirow{3}{*}{\textbf{T}} & \multicolumn{3}{c}{\rule{0pt}{13pt} $\mathbf{\brho Y_{e} = 10^{10}}$} &  & \multicolumn{3}{c}{$\mathbf{\brho Y_{e} = 10^{11}}$} \\[0.4ex]
\cmidrule{2-4}  \cmidrule{6-8} &
\multicolumn{1}{c}{$\blambda_{\bf{QRPA}}$} &
\multicolumn{1}{c}{$\blambda_{\bf{LBA}}$} &
\multicolumn{1}{c}{$\bDelta_{\bf{LBE}}$} & &
\multicolumn{1}{c}{$\blambda_{\bf{QRPA}}$} &
\multicolumn{1}{c}{$\blambda_{\bf{LBA}}$} &
\multicolumn{1}{c}{$\bDelta_{\bf{LBE}}$}
\\
\midrule[1pt]
    1     & 3.78$\times$10$^{-39}$ & 5.61$\times$10$^{-38}$ & -13.86 &       & 1.00$\times$10$^{-100}$ & 1.00$\times$10$^{-100}$ & 0.00 \\
    1.5   & 7.94$\times$10$^{-27}$ & 1.26$\times$10$^{-25}$ & -14.81 &       & 6.78$\times$10$^{-70}$ & 1.77$\times$10$^{-68}$ & -25.06 \\
    2     & 1.45$\times$10$^{-20}$ & 2.42$\times$10$^{-19}$ & -15.75 &       & 7.14$\times$10$^{-53}$ & 1.91$\times$10$^{-51}$ & -25.79 \\
    3     & 3.48$\times$10$^{-14}$ & 6.55$\times$10$^{-13}$ & -17.79 &       & 9.89$\times$10$^{-36}$ & 2.83$\times$10$^{-34}$ & -27.64 \\
    5     & 6.44$\times$10$^{-09}$ & 1.55$\times$10$^{-07}$ & -22.99 &       & 7.28$\times$10$^{-22}$ & 2.41$\times$10$^{-20}$ & -32.11 \\
    10    & 1.00$\times$10$^{-04}$ & 3.27$\times$10$^{-03}$ & -31.58 &       & 3.05$\times$10$^{-11}$ & 1.24$\times$10$^{-09}$ & -39.83 \\
    15    & 4.46$\times$10$^{-03}$ & 1.22$\times$10$^{-01}$ & -26.29 &       & 1.80$\times$10$^{-07}$ & 6.11$\times$10$^{-06}$ & -32.88 \\
    20    & 4.17$\times$10$^{-02}$ & 8.30$\times$10$^{-01}$ & -18.91 &       & 1.91$\times$10$^{-05}$ & 4.78$\times$10$^{-04}$ & -24.06 \\
    25    & 1.81$\times$10$^{-01}$ & 2.75$\times$10$^{+00}$ & -14.21 &       & 3.52$\times$10$^{-04}$ & 6.84$\times$10$^{-03}$ & -18.45 \\
    30    & 5.04$\times$10$^{-01}$ & 6.22$\times$10$^{+00}$ & -11.36 &       & 2.58$\times$10$^{-03}$ & 4.15$\times$10$^{-02}$ & -15.11 \\
\bottomrule
\end{tabular}}
\end{table}

\begin{table}[pt]
\caption{\small The QRPA computed total GT strength (arbitrary units) and centroid values (MeV),  along EC  and BD directions, of  parent states (E$_{x}$) for selected nuclei.  The cutoff excitation energy in daughter states is 20 MeV.}\label{table:table7}
\setlength{\aboverulesep}{0pt} \setlength{\belowrulesep}{0pt} \setlength{\tabcolsep}{15pt} \setlength{\arrayrulewidth}{5pt}
\setlength\heavyrulewidth{2pt} \renewcommand{\arraystretch}{1.15}
{\small
\centering  \hspace{20pt}
\begin{tabular}{ccc|ccc}
\toprule
\multicolumn{3}{|c}{ \textbf{EC} } & \multicolumn{3}{c|}{ \textbf{BD} }\\[0.4ex]
\midrule[2pt]
\midrule[1pt]
\multicolumn{3}{c}{\textbf{$^{78}$Ge}} & \multicolumn{3}{c}{\textbf{$^{58}$Cr}} \\[0.4ex]
\cmidrule{1-3}  \cmidrule{4-6}
\multicolumn{1}{c}{E$_{x}$} &
\multicolumn{1}{c}{$\Sigma B(GT)_{+}$} &
\multicolumn{1}{c}{$\bar{E}_{+}$} &
\multicolumn{1}{c}{E$_{x}$} &
\multicolumn{1}{c}{$\Sigma B(GT)_{-}$} &
\multicolumn{1}{c}{$\bar{E}_{-}$}
\\
\cmidrule{1-3}  \cmidrule{4-6}
    0.00  & 7.07  & 3.06  & 0.00  & 30.39 & 13.68 \\
    2.92  & 11.08 & 3.59  & 3.42  & 30.67 & 15.90 \\
    3.46  & 11.39 & 4.26  & 3.92  & 35.92 & 13.08 \\
    3.69  & 11.92 & 4.42  & 4.27  & 34.64 & 15.88 \\
    4.09  & 11.34 & 4.92  & 4.52  & 39.78 & 13.08 \\
    4.52  & 9.64  & 5.30  & 4.88  & 64.01 & 15.57 \\
    4.75  & 21.77 & 5.40  & 5.14  & 68.15 & 16.07 \\
    5.05  & 21.60 & 5.70  & 5.48  & 68.76 & 15.89 \\
    5.38  & 24.49 & 6.01  & 5.66  & 48.18 & 14.31 \\
    5.66  & 21.51 & 6.29  & 5.98  & 63.47 & 15.94 \\
\midrule[1pt]
\midrule[1pt]
\multicolumn{3}{c}{\textbf{$^{67}$Ni}} & \multicolumn{3}{c}{\textbf{$^{67}$Co}} \\[0.4ex]
\cmidrule{1-3}  \cmidrule{4-6}
\multicolumn{1}{c}{E$_{x}$} &
\multicolumn{1}{c}{$\Sigma B(GT)_{+}$} &
\multicolumn{1}{c}{$\bar{E}_{+}$} &
\multicolumn{1}{c}{E$_{x}$} &
\multicolumn{1}{c}{$\Sigma B(GT)_{-}$} &
\multicolumn{1}{c}{$\bar{E}_{-}$}
\\
\cmidrule{1-3}  \cmidrule{4-6}
    0.00  & 0.74  & 1.78  & 0.00  & 38.73 & 15.49 \\
    0.07  & 0.76  & 3.08  & 0.13  & 39.31 & 15.55 \\
    1.11  & 17.75 & 1.36  & 0.45  & 37.83 & 13.87 \\
    2.24  & 10.61 & 2.27  & 2.19  & 36.46 & 17.54 \\
    3.85  & 15.36 & 6.54  & 2.52  & 41.95 & 17.60 \\
    4.79  & 16.74 & 5.17  & 2.99  & 54.35 & 17.44 \\
    4.98  & 16.77 & 5.82  & 3.36  & 91.16 & 15.87 \\
    5.58  & 17.16 & 5.47  & 3.55  & 163.74 & 14.64 \\
    5.92  & 17.46 & 7.81  & 3.76  & 214.12 & 14.58 \\
    5.98  & 18.41 & 7.04  & 3.94  & 243.04 & 14.59 \\
\midrule[1pt]
\midrule[1pt]
\multicolumn{3}{c}{\textbf{$^{57}$Ni}} & \multicolumn{3}{c}{\textbf{$^{70}$Cu}} \\[0.4ex]
\cmidrule{1-3}  \cmidrule{4-6}
\multicolumn{1}{c}{E$_{x}$} &
\multicolumn{1}{c}{$\Sigma B(GT)_{+}$} &
\multicolumn{1}{c}{$\bar{E}_{+}$} &
\multicolumn{1}{c}{E$_{x}$} &
\multicolumn{1}{c}{$\Sigma B(GT)_{-}$} &
\multicolumn{1}{c}{$\bar{E}_{-}$}
\\
\cmidrule{1-3}  \cmidrule{4-6}
    0.00  & 0.78  & 8.78  & 0.00  & 38.43 & 11.10 \\
    0.04  & 1.63  & 10.62 & 0.15  & 49.19 & 10.48 \\
    0.17  & 1.33  & 8.18  & 0.27  & 50.62 & 10.46 \\
    0.25  & 0.83  & 9.08  & 0.33  & 51.82 & 11.12 \\
    0.29  & 0.88  & 7.36  & 0.45  & 65.61 & 11.00 \\
    2.33  & 1.60  & 11.94 & 0.53  & 59.02 & 11.29 \\
    2.61  & 1.59  & 11.23 & 0.60  & 69.41 & 11.62 \\
    2.66  & 1.70  & 12.07 & 0.72  & 92.62 & 11.66 \\
    2.81  & 2.12  & 11.48 & 0.86  & 77.16 & 11.07 \\
    3.09  & 2.42  & 11.91 & 0.98  & 91.72 & 11.11 \\
\bottomrule
\end{tabular}}
\end{table}

\begin{table}[pt]
	\caption{\small Comparison of our EC rates (This work) on $^{63}$Ni by considering lowest 1, 2, 3 and 5 initial states, and  Local BA rates (with BA) with the corresponding LSSM rates, for selected stellar density ([g\;cm$^{-3}$]) and temperature ([GK]) values. All the rates are given in units of s$^{-1}$.}\label{table:table8}
	\setlength{\aboverulesep}{0pt} \setlength{\belowrulesep}{0pt} \setlength{\tabcolsep}{7pt} \setlength{\arrayrulewidth}{5pt}
	\setlength\heavyrulewidth{1.5pt} \renewcommand{\arraystretch}{1.15}
	
	{\small
		\resizebox{1.15\textwidth}{!}{%
			
			\begin{tabular}{|cccccccccccc|}
				\toprule
				\multicolumn{12}{|c|}{\rule{0pt}{13pt} \textbf{$^{63}$Ni} } \\[0.4ex]
				\midrule[2pt]
				\midrule[1pt]
				
				&\multirow{1}[2]{*}{} &  \multicolumn{2}{c}{state 1} & \multicolumn{2}{c}{state 2} & \multicolumn{2}{c}{state 3} & \multicolumn{2}{c}{state 5} & \multicolumn{2}{c|}{With BA} \\
				
				\cmidrule{3-12}
				\textbf{T} &  $\mathbf{\brho Y_{e}}$ & This work & LSSM  & This work & LSSM  & This work & LSSM  & This work & LSSM & This work & LSSM  \\
				\midrule[1pt]
				3 & 10$^{7}$ & 3.59$\times$10$^{-11}$ & 2.40$\times$10$^{-10}$ & 2.25$\times$10$^{-10}$ & 3.47$\times$10$^{-09}$ & 2.96$\times$10$^{-10}$ & 2.57$\times$10$^{-09}$ & 1.75$\times$10$^{-09}$ & 2.40$\times$10$^{-09}$ & 2.21$\times$10$^{-09}$ & 3.47$\times$10$^{-09}$
				\\
				4 & 10$^{8}$ & 2.32$\times$10$^{-07}$ & 1.48$\times$10$^{-06}$ & 8.99$\times$10$^{-07}$ & 7.59$\times$10$^{-06}$ & 1.12$\times$10$^{-06}$ & 5.62$\times$10$^{-06}$ & 3.84$\times$10$^{-06}$ & 5.13$\times$10$^{-06}$ & 4.99$\times$10$^{-06}$ & 6.61$\times$10$^{-06}$
				\\
				5 & 10$^{9}$ & 1.79$\times$10$^{-03}$ & 1.17$\times$10$^{-02}$ & 4.73$\times$10$^{-03}$ & 2.45$\times$10$^{-02}$ & 5.61$\times$10$^{-03}$ & 1.91$\times$10$^{-02}$ & 9.27$\times$10$^{-03}$ & 1.70$\times$10$^{-02}$ & 1.11$\times$10$^{-02}$ & 1.82$\times$10$^{-02}$
				\\
				7 & 10$^{10}$ & 2.14$\times$10$^{+01}$ & 6.92$\times$10$^{+01}$ & 3.36$\times$10$^{+01}$ & 6.31$\times$10$^{+01}$ & 4.37$\times$10$^{+01}$ & 6.03$\times$10$^{+01}$ & 5.25$\times$10$^{+01}$ & 5.50$\times$10$^{+01}$ & 6.47$\times$10$^{+01}$ & 6.03$\times$10$^{+01}$ \\
				\bottomrule
	\end{tabular}}}
	
\end{table}

\begin{table}[pt]
	\caption{\small Comparison of our BD rates (This work) of $^{63}$Co by considering lowest 1, 2 and 4 initial states, and Local BA rates (with BA) with the corresponding LSSM rates, for selected stellar density ([g\;cm$^{-3}$]) and temperature ([GK]) values. All the rates are given in units of s$^{-1}$.}\label{table:table9}
	\setlength{\aboverulesep}{0pt} \setlength{\belowrulesep}{0pt} \setlength{\tabcolsep}{7pt} \setlength{\arrayrulewidth}{5pt}
	\setlength\heavyrulewidth{2pt} \renewcommand{\arraystretch}{1.15}
	{\small
		\resizebox{1.15\textwidth}{!}{%
			\begin{tabular}{|cccccccccc|}
				\toprule
				\multicolumn{10}{|c|}{\rule{0pt}{13pt} \textbf{$^{63}$Co} } \\[0.4ex]
				\midrule[2pt]
				\midrule[1pt]
				
				&\multirow{1}[2]{*}{} &  \multicolumn{2}{c}{state 1} & \multicolumn{2}{c}{state 2} & \multicolumn{2}{c}{state 4} & \multicolumn{2}{c|}{With BA} \\
				
				\cmidrule{3-10}
				\textbf{T} &  $\mathbf{\brho Y_{e}}$ & This work & LSSM  & This work & LSSM  & This work & LSSM & This work & LSSM    \\
				\midrule[1pt]
				3     & 10$^{7}$     & 8.67$\times$10$^{-03}$ & 2.29$\times$10$^{-02}$ & 4.16$\times$10$^{-02}$ & 2.40$\times$10$^{-02}$ & 6.07$\times$10$^{-02}$ & 2.45$\times$10$^{-02}$ & 6.08$\times$10$^{-02}$ & 2.51$\times$10$^{-02}$  \\
				4    & 10$^{8}$     & 3.06$\times$10$^{-03}$ & 1.07$\times$10$^{-02}$ & 1.94$\times$10$^{-02}$ & 1.26$\times$10$^{-02}$ & 3.56$\times$10$^{-02}$ & 1.29$\times$10$^{-02}$ & 3.73$\times$10$^{-02}$ & 1.70$\times$10$^{-02}$ \\
				5     & 10$^{9}$     & 1.61$\times$10$^{-05}$ & 9.33$\times$10$^{-05}$ & 1.88$\times$10$^{-04}$ & 2.24$\times$10$^{-04}$ & 5.13$\times$10$^{-04}$ & 3.02$\times$10$^{-04}$ & 2.09$\times$10$^{-03}$ & 2.82$\times$10$^{-03}$ \\
				7     & 10$^{10}$    & 3.40$\times$10$^{-09}$ & 1.95$\times$10$^{-08}$ & 3.24$\times$10$^{-08}$ & 5.01$\times$10$^{-08}$ & 9.93$\times$10$^{-08}$ & 7.59$\times$10$^{-08}$ & 1.66$\times$10$^{-05}$ & 3.72$\times$10$^{-06}$ \\
				
				\bottomrule
	\end{tabular}}}
	
\end{table}

\end{document}